\documentclass[10pt,conference]{IEEEtran}
\usepackage{tikz}
\usepackage{amsmath}
\usepackage{graphicx}
\usepackage{xspace}
\usepackage{subcaption}
\usepackage{tabularx}
\usepackage{harveyballs}
\usepackage{fancyhdr}
\usepackage[utf8]{inputenc}
\usepackage{longtable}
\usepackage[hyphens]{url}
\fancypagestyle{firstpage}{%
    \setlength{\headheight}{0.0in}
    \setlength{\headsep}{0.15in}
    \chead{Confidential draft: To appear in IEEE Symposium on Security and Privacy, 2021}
}
\begin{document}
%-------------------------------------------------------------------------------

%don't want date printed
\date{}

\title {Doing good by fighting fraud: Ethical anti-fraud systems for mobile payments}

%\author{
%  {\rm Anonymous submission}}

\author{
    \IEEEauthorblockN{
        Zainul Abi Din\IEEEauthorrefmark{1}, Hari Venugopalan\IEEEauthorrefmark{1}, Henry Lin\IEEEauthorrefmark{2}, Adam Wushensky\IEEEauthorrefmark{2}, Steven Liu\IEEEauthorrefmark{2},
        Samuel T. King\IEEEauthorrefmark{1}\IEEEauthorrefmark{2}
    }
    \IEEEauthorblockA{\IEEEauthorrefmark{1} University of California, Davis}
    \IEEEauthorblockA{\IEEEauthorrefmark{2} Bouncer Technologies}
}

\newcommand{\totalscans}{1,103,524\xspace} % From open source
\newcommand{\totalscansabr}{1.1 million\xspace} % From open source
\newcommand{\lowendacclow}{57.70\%\xspace} % Android Ovefeat
\newcommand{\lowendacchigh}{89.08\%\xspace} % Android DD V4
\newcommand{\lowframeratepercentage}{45.27\%\xspace} % percentage of devices below 1 fps
\newcommand{\overallscans}{50 million\xspace} % includes everything
\newcommand{\totalapps}{496\xspace}
\newcommand{\boxerfail}{53.28\%\xspace}
\newcommand{\boxertotal}{3,505,184\xspace}
\newcommand{\daredeviltotal}{1,580,260\xspace}

\maketitle
\newcommand{\system}{Daredevil\xspace}
\newcommand{\othersystem}{Boxer\xspace}

% Remove if you don't want the confidential header
%\thispagestyle{firstpage}
% ----- Confidential header ends

\begin{abstract}

App builders commonly use security challenges, a form of step-up
authentication, to add security to their apps. However, the ethical
implications of this type of architecture has not been studied
previously.
  
In this paper, we present a large-scale measurement study of running
an existing anti-fraud security challenge, Boxer, in real apps running
on mobile devices. We find that although Boxer does work well overall,
it is unable to scan effectively on devices that run its machine
learning models at less than one frame per second (FPS), blocking
users who use inexpensive devices.

With the insights from our study, we design \system, a new anti-fraud
system for scanning payment cards that works well across the broad range
of performance characteristics and hardware configurations found on
modern mobile devices. \system reduces the number of devices that run
at less than one FPS by an order of magnitude compared to Boxer,
providing a more equitable system for fighting fraud.

In total, we collect data from 5,085,444 real devices spread across
496 real apps running production software and interacting with real
users.

\end{abstract}

\section{Introduction}

Smartphones and apps are ubiquitous, with billions of daily users and
over 5 million apps available for everything from dating and travel to
payments and food deliveries.  Unfortunately, smartphones and apps
have also ushered in a new generation of attacks \cite{uber_reselling,
  airbnb_money_launder, uber_fraud_article}, forcing app builders to
design and implement user-centric security measures, or
\emph{challenges}, in their apps \cite{lyft_blog,
  airbnb_false_positive, netverify}. Examples of this new style of
verification include Apple's FaceID where they use face biometrics to
authenticate a user \cite{face_id}, Uber's credit-card scanning where
they ask users to scan their card to prove that they possess it
\cite{uber_verify, ramasamy18}, Coinbase's ID verification where they
ask users to scan an ID to prove who they are in the real world
\cite{coinbase_id}, and Lime's Access program that allows people of a
low socioeconomic status to scan IDs and utility bills to prove that
they qualify for discounted rental fees \cite{lime_access}.

Challenges have the potential to skirt the difficult ethical issues
that apps face with security decisions in their apps. In a typical
app, the app will have an algorithm that predicts whether a user or a
transaction is suspicious. These algorithms could potentially rely on
features that unfairly influence its decision, such as a zip code. To
reduce the impact of mistakes by their algorithms, apps can use
user-centric security measures in lieu of suspending users or blocking
transactions. This technique allows users that the algorithm blocks
incorrectly to verify themselves or their payment methods
automatically. Thus, even if their algorithm has bias
\cite{mlbiasequalizedodds}, challenges provide an avenue for making
sure that everyone can access the app.

Unfortunately, challenges open a new set of ethical
conundrums. Apps that want to respect end-user privacy and run their
challenges via compute intensive machine learning models on the device
will have to cope with the 1-4 orders of magnitude difference in
capabilities on the devices that they will see in practice (Section
\ref{sec:measurement_study}).  Apps that opt for predictable ML
performance by streaming data to a server and running their ML there
will have to deal with a 1000x difference in bandwidth between 3G and
5G networks \cite{verizon_3g_5g}, and the people who use it may have
to pay for that bandwidth directly. Security challenges must deal with
these subtleties of practical deployments or else they will block
users unethically.

The most dangerous aspect of the ethical implications inherent with
security challenges is that they solve an app's business problem but
have the potential to still make compromises on users of a low
socioeconomic status. One example of this tradeoff is with Lime's Access
program \cite{lime_access}. Lime allows users from low-income households to 
get reduced rates with Lime rentals by proving that they qualify for the 
program by scanning welfare documents or utility bills. These documents 
contain personal information that typical Lime users do not have to provide,
and Lime does not process these documents themselves, they use a third
party for this service \cite{lime_fountain}. Just to be clear, we, as
proponents of this program applaud Lime for implementing it, but Lime
forces users to give up privacy to qualify. It would be
better if they could prove that these documents are genuine without
needing to send sensitive information to a third-party server.

A second example is Boxer \cite{Din20}, a system presented at Usenix
Security 2020 for scanning credit cards to prove that the user
possesses the genuine physical card. Boxer uses client-side machine
learning to verify the credit card. However, based on our measurement
study of Boxer's open-source card scanner (Section
\ref{sec:measurement_study}), Boxer fails on 68.13\% of Android
devices that run its ML at less than one FPS. Slower ML inference
corresponds to lower frame rates and thus, fewer inputs that the
system processes for verification.  Like Lime's Access program, apps
that use Boxer solve their business problem -- only 4.19\% of the
total devices that we measure are Android devices that run Boxer's ML
at less than one FPS. By using Boxer, apps recover most of the people
that their security systems flag incorrectly. However, by blocking
devices that are unable to run their ML models fast enough, they run
the risk of denying access to at-risk populations simply because they
have an inexpensive device.

The inability to run challenges on resource-constrained devices
introduces a new bias that the existing formulations of machine
learning fairness \cite{mlbias1, mlbias2, mlbiasequalizedodds, causal,
  puttingfairnessintopractice} are ill-equipped to solve. Existing
formulations of machine learning fairness modify either the decision
engine or the feature set corresponding to an individual of a
protected group to ensure that protected attributes (e.g., race) do
not affect the outcome.  However, being unable to run models on
resource-constrained devices robs the decision engine of the inputs it
needs to make a decision in the first place. Although the decision
engine could randomly pass individuals whose devices are low end or
randomly block otherwise good users to provide a notion of fairness,
both degrade the performance of the overall system since they weaken
their ability to distinguish between legitimate and fraudulent users. No
algorithmic or theoretical notion of fairness can account for this
lack of data.

Our position is that ethical security challenges should run client
side, support complex machine learning (if needed), and run effectively on
resource-constrained devices.  In this paper, we present \system, a system
for running complex client-side ML models for security on the full
range of devices one is likely to see in practice today. \system's
design includes decomposing machine learning tasks for redundancy and
efficiency, streamlining individual tasks for improved performance,
and exploiting task and data parallelism.

We demonstrate \system by designing and implementing a new credit card
scanning and verification system. Card scanners use complex machine
learning models and hundreds of apps use them in practice
\cite{Din20}, which make them a good candidate for \system. We deploy
\system to real apps and demonstrate how it provides access to a wide
range of devices. Through our deployment, we run \system on over
\daredeviltotal devices from real users and show how \system both
enables resource-constrained hardware to run security ML models
effectively, and it improves the end-to-end success rates on
well-provisioned hardware with support for fast ML models.
%suggesting that the
%principles we apply for \system are general.

Our contributions are:
\begin{itemize}
    %\item We introduce a new machine learning bias inherent in security
    %  challenges for defending against financial fraud, that can
    %  result in disproportionately adverse outcomes for certain
    %  groups of individuals, improving the security and user experience
    %  of one group at the expense of altogether blocking the other. 
    
    \item We present the first large-scale in-field study of on-device
      deep learning for security.  Our measurements focus on Boxer, a
      system for scanning credit cards, where we demonstrate that due
      to the degree of hardware diversity, deep-learning-based
      security challenges have the potential of being unethical
      despite solving the apps' business problem.
    \item We uncover insights from our measurement study such as
      critical reasons for failure cases, key system metrics, and
      mitigation strategies that developers should consider when
      designing a client-side machine learning pipeline.
    \item Equipped with the insights from our measurement study, we
      design, implement, and deploy \system, which empowers card
      scanning and verification to run on a wide range of devices.
\end{itemize}

\section{Background: Card-not-present credit card fraud and card scanning}
\label{sec:background}

Fraudsters acquire stolen credit card information and use it to make
purchases online, without possessing the actual physical card. This
is known as card-not-present credit card fraud. When the real owner 
of the card notices a suspicious charge on their credit card statement, 
they report it to the credit card company. Upon investigating the
transaction, if the credit card company finds the transaction to be
fraudulent, they will issue a chargeback to the app.  The app will
have to pay back the money to the real owner of the card, and an
additional dispute fee to the credit card company
~\cite{stripe_disputes}.  This protects the owner of the credit card
and puts the responsibility of curbing card-not-present credit card
fraud on the app.

Recently, researchers propose Boxer~\cite{Din20}, a mobile SDK and
server that app builders integrate with apps to prevent
card-not-present credit card fraud.  Boxer shows how to scan the
number side of a card and verify that it is genuine. Boxer casts card
verification as a machine learning problem that it divides into three
main parts: optical character recognition (OCR), fake media detection
(implemented via screen detection in their paper), and card tampering
detection (called a \emph{Bank Identification Number or BIN  consistency check} in their paper).  OCR
pulls the card number, expiration, and legal name off the card. Screen
detection detects when a user scans a card from a screen instead of
using a physical card. Card tampering detection finds prominent objects
on the card, like the bank logo, and correlates this information with
the OCR prediction to confirm that these objects are consistent with
the type of card that they expect. For instance, if OCR detects a BIN
(first six digits of the card number) of a Chase Visa payment card but
the card tampering detection detects a Bank of America logo or Mastercard
logo, Boxer flags this scan as fraudulent.

However, Boxer falls short in the following ways: First, OCR, as their
first line of defense, stops the vast majority of fraudsters as per
their evaluation. However, as we describe in our measurement study
(Section \ref{sec:measurement_study}), Boxer's OCR under performs on 
low-end devices.
Second, Boxer's fraud checks only
scan the number side of the card. However, newer card designs in the wild
contain visual design elements on either side of the card, so by scanning
only the number side of the card Boxer misses out on key information. Third,
Boxer only flags cards scanned off screens and not other fake media.

In the remainder of the paper, we first describe our measurement study
of Boxer (Section \ref{sec:measurement_study}). We then describe the
design of \system, a new credit card scanning and verification system
to improve upon Boxer (Sections \ref{sec:overview},
\ref{sec:design}). This is followed by a detailed evaluation of
\system(Section \ref{sec:evaluation}).

\newcommand{\boxertotalapps}{496\xspace}
\newcommand{\boxertotalandroid}{329,272\xspace}
\newcommand{\boxertotalandroidtype}{611\xspace}

\newcommand{\boxertotalios}{3,175,912\xspace}
\newcommand{\boxertotaliostype}{27\xspace}

\newcommand{\boxertotalfailure}{537,359\xspace}
\newcommand{\boxeriosfailure}{361,924\xspace}
\newcommand{\boxerandroidfailure}{175,435\xspace}

\newcommand{\samsungtype}{281\xspace}
\newcommand{\samsungtotal}{168,658\xspace}

\newcommand{\huaweitype}{91\xspace}
\newcommand{\huaweitotal}{49,329\xspace}

\newcommand{\xiaomitype}{64\xspace}
\newcommand{\xiaomitotal}{80,351\xspace}

\newcommand{\lgtype}{63\xspace}
\newcommand{\lgtotal}{5,464\xspace}

\newcommand{\googletype}{11\xspace}
\newcommand{\googletotal}{2,939\xspace}

\newcommand{\motorolatype}{27\xspace}
\newcommand{\motorolatotal}{2,501\xspace}

\newcommand{\oneplustype}{18\xspace}
\newcommand{\oneplustotal}{2,560\xspace}

\newcommand{\tailtype}{56\xspace}
\newcommand{\totaltail}{17,470\xspace}
\newcommand{\tailvendors}{23\xspace}

\section{Measurement study}
\label{sec:measurement_study}

In this section, we present the first large-scale measurement study of
a security challenge using deep learning on mobile devices. We study the practical characteristics and limitations of
credit card scanning using real apps running on end-user devices with
real people and credit cards, and all the idiosyncrasies inherent
in large-scale software with live deployments.

We believe this study is the first of its kind and has implications
for deep learning engineers, app developers, and hardware vendors.
The closest to our study in terms of scale is presented by Ignatov
et. al \cite{IgnatovAI}, however, their study is limited to only
10,000 Android devices and runs pre-defined images through pre-trained
models loaded on each device to benchmark the hardware.

Ours is an in-field correlation study and represents a realistic usage
scenario for end-users since we benchmark the usage of a deep learning
driven application where the user, the phone sensor, image processing,
ambient lighting, device surface temperature, the compute capability of
the device and other production variables determine the performance of
the system. In our study, we protect end-user privacy by limiting the
amount and nature of the statistics that we record, the metrics have
enough fidelity that they inform our end-to-end design (Section
\ref{sec:design}), resulting in significant improvements in the wild
(Section \ref{sec:evaluation}).

Our university's IRB board reviewed our study and ruled it to be
exempt from IRB. 

\subsection{Measurement study goals and questions}

Our high-level goal is to understand the practical performance and
limitations of camera-based mobile security challenges in real-world
conditions. We perform our study using Boxer, a widely deployed credit
card scanning system and measure its success rate as the primary
metric for success. To understand the performance and limitations, we
focus our correlation study on three primary questions.

\emph{How does the speed of ML predictions influence end-to-end
  metrics for success?} The research community and industry have put a
heavy emphasis on performance for ML predictions through machine
learning models designed specifically for mobile devices
\cite{mobilenet, squeezenet, mobilenetv2, mobilenetv3} and hardware
support for fast inference \cite{coreml, edgetpu}. We measure the
impact of these efforts on high-level metrics for success.

\emph{How widely do the ML capabilities on modern phones vary in the
  field?} We measure the range of ML capabilities one is likely to see
in practice. By understanding the range of capabilities, one can
anticipate the performance differences for security challenges in
realistic settings. Also, we quantify the number of devices that are
unable to run Boxer ML effectively, which for a security check blocks
the user.

\emph{How long are people willing to wait when they try to scan
  documents with their phone?} As there are many forms of scanning
documents that apps use for security checks, understanding how long
people are willing to wait as they try to scan informs the overall
design of a security check. Security check designers will know how
long they have to capture relevant information before someone gives
up.

\subsection{Measurement Platform}

To measure Boxer's performance, we instrumented Boxer's open-source SDK and made it available to 
third-party app developers. We then measured the success rate for the users of their
live production apps.  We present results from anonymous statistics
sent by \boxertotalapps apps that deployed and ran the instrumented library 
from July 2019 to late November 2020. 

\subsection{Testbed}
Our instrumented Android SDK ran on a total of \boxertotalandroid Android devices spanning a total of
\boxertotalandroidtype Android device types. This included \samsungtotal Samsung devices
spanning \samsungtype Samsung device types,  \huaweitotal Huawei devices spanning \huaweitype
Huawei device types, \xiaomitotal Xiaomi devices spanning \xiaomitype
Xiaomi device types, \lgtotal LG devices spanning \lgtype
LG device types, \googletotal Google devices spanning \googletype
Google device types, \motorolatotal Motorola devices spanning \motorolatype
Motorola device types, \oneplustotal OnePlus devices spanning \oneplustype
OnePlus device types and tail of \totaltail devices, spanning \tailtype device types
and \tailvendors vendors.
Our instrumented iOS SDK ran on a total of \boxertotalios iOS devices spanning a total of
\boxertotaliostype iOS device types. 

\subsection{Task}
\label{sec:measurement_study:task}
Our instrumented SDK prompts users to scan their credit cards. When invoked, it starts the camera and prompts users to place their card in the center of the viewport. The OCR processes the frames obtained
from the camera and attempts to extract the card number and the expiry from the card. Upon success, 
the card number and the expiry are displayed to the user and the SDK sends the scan statistics
to our server. In case, the OCR is unable to extract the number, the flow doesn't time-out, instead
we let the user cancel the scan which provides us an additional user-level metric that can guide 
a new design.

Boxer uses a two-stage OCR similar to
Deep Text Spotter~\cite{Busta_2017_ICCV}, consisting of a
detection phase to detect groups of digits in credit card images and a
recognition phase to extract individual digits from these groups. 
%The input image
%from the camera is processed and sent to the detection model, which outputs a set of proposals.
%These proposals are then passed to the second model that outputs the digits contained in the
%proposals. Afterwards, a set of post-processing methods are run to aggregate the information.

Both models use a modified version of MobileNet \cite{mobilenet}, where the detection model
occupies 1.75MB and the recognition model occupies 1.19MB on disk.

The detection model processes an input image of size 480x302x3 and generates
a total of 1,734 proposals.
It has a total of 910,379 parameters of which 901,379 are trainable. There 
are 16 2D convolution blocks and 10 Depthwise convolution blocks, each followed by a 
batch-norm and an activation layer.

The recognition model processes input images of size 36x80x3, each 
corresponding to the proposal generated by the detection model and
generates a feature map of 17x11 for each proposal. It has a total
of 618,540 parameters of which 611,754 are trainable. There are
14 2D convolution blocks and 8 Depthwise convolution blocks, each followed
by a batch-norm and an activation layer.

For inference, the iOS SDK uses the vendor specific CoreML
which runs the models on the CPU, the GPU and the Neural engine
depending upon the availability and usage at any time.
The Android SDK uses a generic interpreter TFlite, where the
inference primarily runs on the CPU. Boxer's models 
are quantized using 16-bit floating point weights.

\subsection{Results} 

\subsubsection{Key Performance Metrics}
\textbf{Success rate:} We define success rate as the ratio of the
number of users where the scanner successfully extracted the card
number to the total number of users using the scanner.

\textbf{Frame rate:} We define the frame rate as the number of frames
from the camera processed by the OCR pipeline (detection and
recognition) per second.

%\textbf{Low-end device:} We define low-end devices as devices that run
%the Boxer ML at frame rate of less than 1 frame-per-second (FPS).

\begin{figure}[t]
    \centering
    \includegraphics[width=\columnwidth]{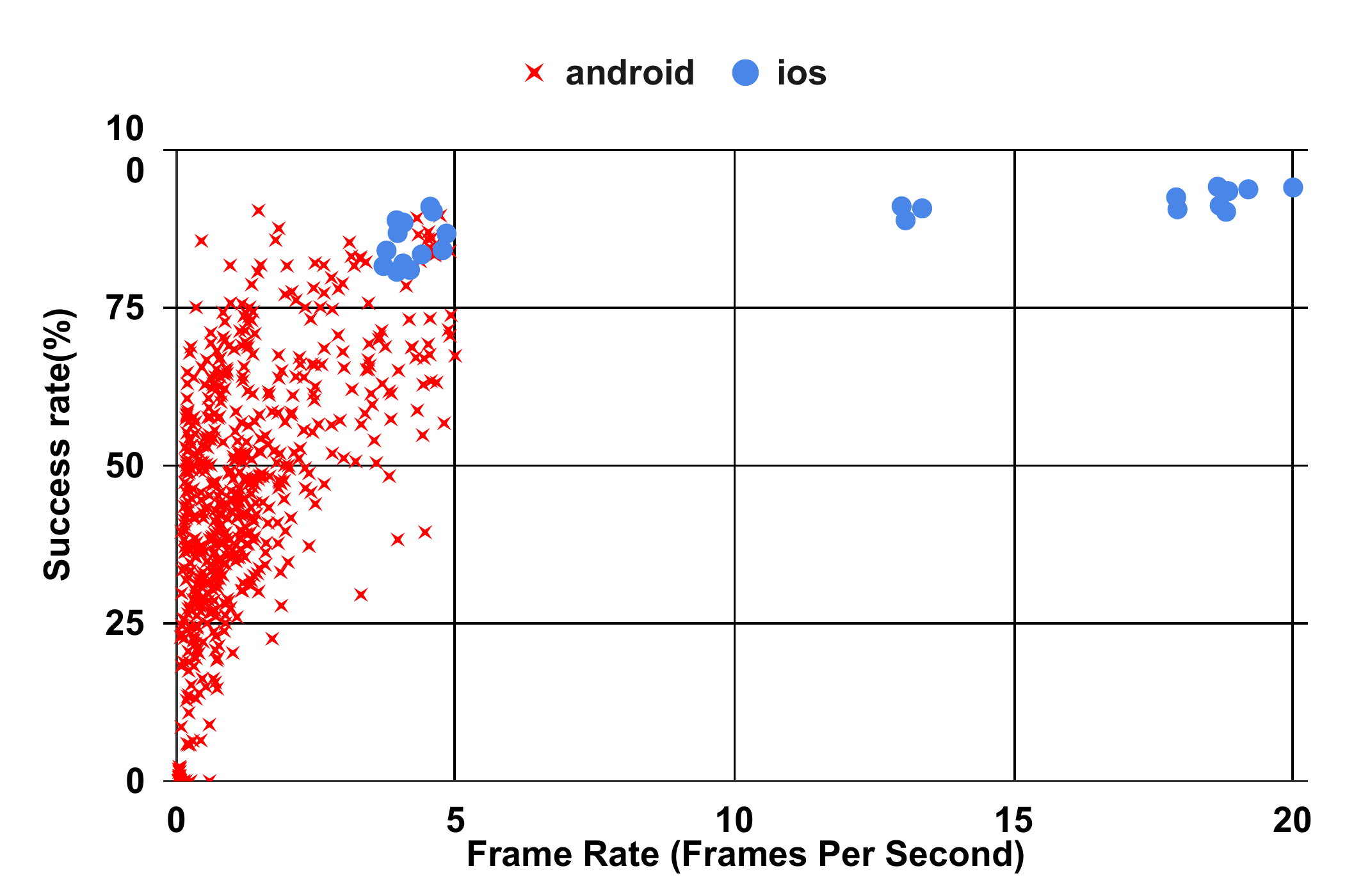}
    \caption{Boxer OCR success rate vs frame rate on Android and iOS. 
      Each point is the average success rate and
      frame rate for a specific device type. This figure shows that
      when using the same machine learning model, end-to-end success
      rates drop off as the frame rate declines. We also see the same
      model and system architecture exhibit different performance 
      characteristics on Android and iOS.}
    \label{fig:boxer_ocr_visual}
    \hrulefill
\end{figure}

\begin{figure}[t]
  \centering
  \small
  \begin{tabular}{|l|c|c|c|}
    \hline
    {\bf Android FPS} & {\bf Count} & {\bf Success rate} \\ \hline \hline
    $<$ 1 FPS & 146,890 (44.61\%) & 31.87\% \\ \hline
    1$-$2 FPS & 97,798 (29.70\%) & 49.97\% \\ \hline
    $>=$ 2 FPS & 84,584 (25.68\%) & 68.72\% \\ \hline
  \end{tabular}
  \caption{Success rates for Android devices running Boxer by the frame rate. We see that a significant portion of devices operate at frame rates less than 1 FPS.}
  \label{fig:boxer_android_frame_rate}
  \hrulefill
\end{figure}

Figure \ref{fig:boxer_ocr_visual} shows the variation in success rate
against the frame rate for different devices.  We 
omit iPhone 6 and below devices from our deployment since
Boxer does not support them.

Data from Figure \ref{fig:boxer_ocr_visual} and Figure \ref{fig:boxer_android_frame_rate} suggests:
\begin{itemize}
    \item Both the frame rate and the success rate are higher on iOS
      than on Android when using the same machine learning models and
      same system architecture.
    \item Boxer is ineffective on Android devices when the frame rate
      is less than 1 FPS. These devices make up 44.61\% of the Android devices in our 
      study and achieve a success rate of 31.87\% compared to 49.97\% 
      for devices that run at 1-2 FPS and 68.72\% for devices that run at 2 FPS 
      or higher. 
\end{itemize}

\begin{figure}[t]
  \centering
  \small
  \begin{tabular}{|l|c|c|c|c|}
    \hline
    {\bf Platform} & {\bf Count} & {\bf Avg Success} & {\bf Avg} & {\bf Avg}\\
    {\bf        } & {\bf      } & {\bf Rate} & {\bf FPS} & {\bf Duration (s)}\\ \hline \hline
    Android  & \boxertotalandroid  & 46.72\%  & 1.303 & 14.45 \\ \hline
    iOS  & \boxertotalios & 88.60\%  & 10.00 & 10.02 \\ \hline
  \end{tabular}
  \caption{Aggregate results of Boxer on Android and iOS.}
  \label{fig:boxer_android_vs_ios}
  \hrulefill
\end{figure}

Figure \ref{fig:boxer_android_vs_ios} shows aggregate results for iOS and Android.
While the success rate for iOS is 88.60\%, the success rate for Android is much lower at 46.72\%.
%The detailed per-device breakdown is shown in the Appendix.

\begin{figure}[t]
  \centering
  \small
  \begin{tabular}{|l|c|c|c|}
    \hline
    {\bf Platform} & {\bf Count} & {\bf Avg} & {\bf Avg}\\
    {\bf        } & {\bf      } & {\bf FPS} & {\bf Duration (s)}\\ \hline \hline
    Android  & \boxerandroidfailure  & 1.00  & 16.20 \\ \hline
    iOS  & \boxeriosfailure & 9.28  & 20.73  \\ \hline
  \end{tabular}
  \caption{Failure cases of Boxer on Android and iOS.}
  \label{fig:boxer_failure_android_and_ios}
  \hrulefill
\end{figure}

\subsubsection{Further analysis of failure cases}
We measure how long users are willing to wait to scan their cards by
measuring how long people scan for when they are unsuccessful in
scanning their card. Measuring the time that people are willing to
wait while scanning informs our decisions when designing the system
and trading off scan times vs accuracy and fraud signal fidelity.

From our real-world deployment of \boxertotal scans, we observed \boxertotalfailure
failed attempts where users gave up on trying to scan their card. We
aggregate the duration of these scans on iOS and Android to report
that Android users waited an average of 16.20s and iOS users waited an
average of 20.73s to scan their cards before giving up (Figure \ref{fig:boxer_failure_android_and_ios}).

\subsection{Context for the results}
For its fraud challenge, Boxer uses OCR to verify the card number that
the app has on record for any user. Thus, anyone who is unable to scan
their number will be unable to pass the fraud challenge. Additionally,
OCR is the first model in the Boxer pipeline and is used to extract
data like the first six digits (BIN), which is then correlated with
other features like the credit card design to determine fraud.
However, if the first model in the pipeline fails to run, the device
is implicitly denied the service.

%It can be seen from the results that devices that run at a rate lower 
%than one FPS are unable to scan cards despite using the same machine 
%learning models and the same system architecture. 

Boxer solves the business problem that its designers intended to solve, since it 
runs OCR successfully on 84.7\% of the devices overall. However, the
success rate on devices that run at a rate of less than 1 FPS is mere 31.87\%, 
and these devices make up 44.61\% of the Android devices we measure,
introducing a potential ethical conundrum by blocking users solely because they
have an inexpensive device.
% put here so that the two column figure shows up in the right place
\begin{figure*}[t]
\centering
\includegraphics[width=1.0\textwidth]{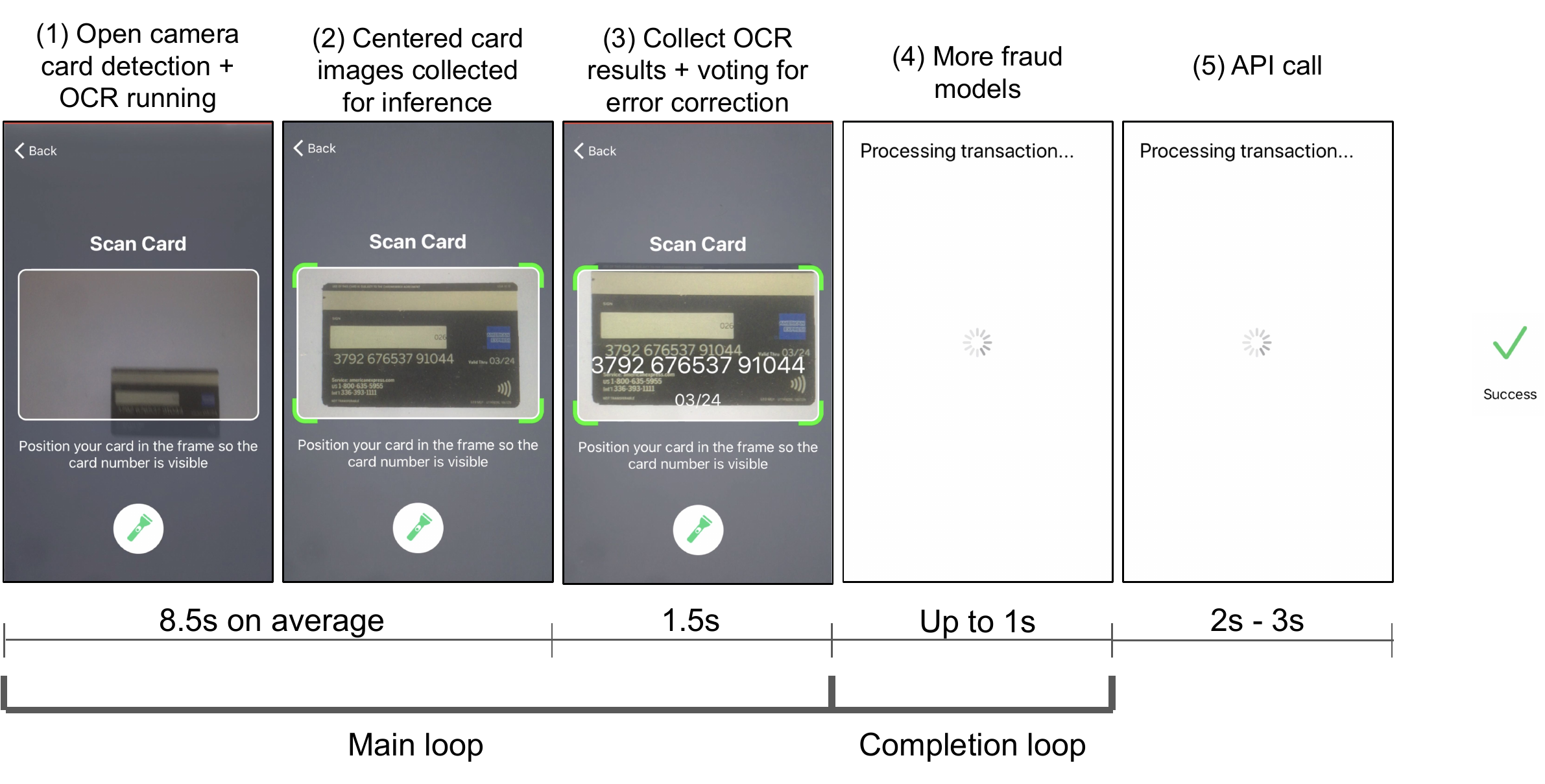}
\caption{\system scanning one side of a card from a user's
  perspective.}
\label{fig:user_flow}
\hrulefill
\end{figure*}

\section{Overview}
\label{sec:overview}
In this paper, we introduce \system, a new system that we design
and implement to realize ethical deep learning powered user-centric
security challenges, with the goal of providing equal access to all users.
Although we built \system to prevent card-not-present credit card fraud, 
the insights gained from \system can also be applied to design other
end-user security challenges.

To provide equal access, \system must be fast, even on resource-constrained
devices that lack hardware acceleration for machine learning, \system must
respect end user privacy, and \system must be accurate to avoid
incorrectly flagging otherwise good users as being fraudulent.

Our work on \system builds off recent work from Din \emph{et al.}
\cite{Din20} that shows how to scan the number side of a card and
verify that it is genuine using a system called Boxer, as described
in Section \ref{sec:background}.

We demonstrate the design of an ethical fraud challenge by improving Boxer 
in the following ways:
\begin{itemize}
    \item  We design a new fast and efficient OCR that also runs well
    on resource-constrained devices.
    \item  We propose a machine learning pipeline that
    combines the different models to provide efficiency and redundancy.
    \item We introduce a new card detection model that operates in concert with 
    card tampering detection to scan both sides of the card.
\end{itemize}

\subsection{Threat model}
In our threat model, our goal is to reduce financial fraud while
ensuring that all users can pass our challenge. Our focus is on
challenges that apps can use to verify that people possess a genuine
credit card.

We assume that the attacker has stolen credit card credentials (e.g.,
the card number and billing ZIP code), but does \emph{not} possess the
real credit card.

Our machine learning models run client side, where \system processes
credit card images on the device before passing a distilled summary of
the machine learning output to our server, where we make the ultimate
decision about if a scan is genuine. As our models run client side, we
are susceptible to attackers who tamper with the app, the video
stream, or our machine learning models. Although we do have some
measures in place to assess the integrity of our client-side software
(e.g., DeviceCheck on iOS and SafetyNet on Android), we recognize that
this type of assurance is still an ongoing arms race between app
builders, device manufacturers, and attackers. Our design favors
end-user privacy even though it does open us up to client-side
attacks.

%To pass our challenge, users need to run our machine learning
%models. For attackers, this eliminates attacks where they try to
%degrade the performance of a device to evade our defenses. However,
%for good users, this architecture sets a high bar for \system because
%we strive to build a challenge that all people can pass, even if they
%use an resource-constrained device.

\subsection{Architecture}

To scan cards and verify that they are genuine, \system asks users to
scan the front of their card and the back. This makes \system flexible
to verify a wide range of card designs where meaningful information can be 
on either side of the card. Scanning both sides also provides more data for 
us to detect signs of tampering than if we 
scan only a single side. Our checks inspect individual card sides to 
ensure that they are genuine, as well as combining information from both 
sides to make sure that it is consistent.

However, scanning both sides of the card complicates the machine
learning aspects of verifying a card. First, credit cards are
free to print design elements on either side. Second, users
(and some of the authors) are unaware of which side of the card is the
front versus the back. Therefore, \system must be flexible enough to pull
out the appropriate information to detect fraud dynamically and adapt
automatically to scan the appropriate side of the card for each
scan. The net result is that to verify cards \system must run more
machine learning models than it would if it were just scanning a
single side of the card.

Figure \ref{fig:user_flow} shows this overall process from a user's
perspective. First, the user (1) opens the flow, which starts the
camera. Then (2) when they put the card in the center of the viewport,
we update the user interface to give them feedback. In parallel, (3)
the card detection and the OCR models run and we display the details
that the OCR extracts from the card. After the first successful OCR
prediction we continue running the card detection and OCR models for
1.5s and collect additional predictions about the OCR details to vote
and correct any mispredictions. After the error correction process
completes, (4) we run the fake media detection and card tampering
detection models on a subset of the images that we process for up to
1s, before (5) making an API call to our server to judge if the scan
included a genuine physical card. This API call includes the output of
our client-side machine learning models and our server-side logic
implements rules to make a final overall decision about the validity
of a scan.

\begin{figure}[t]
\centering
\includegraphics[width=\columnwidth]{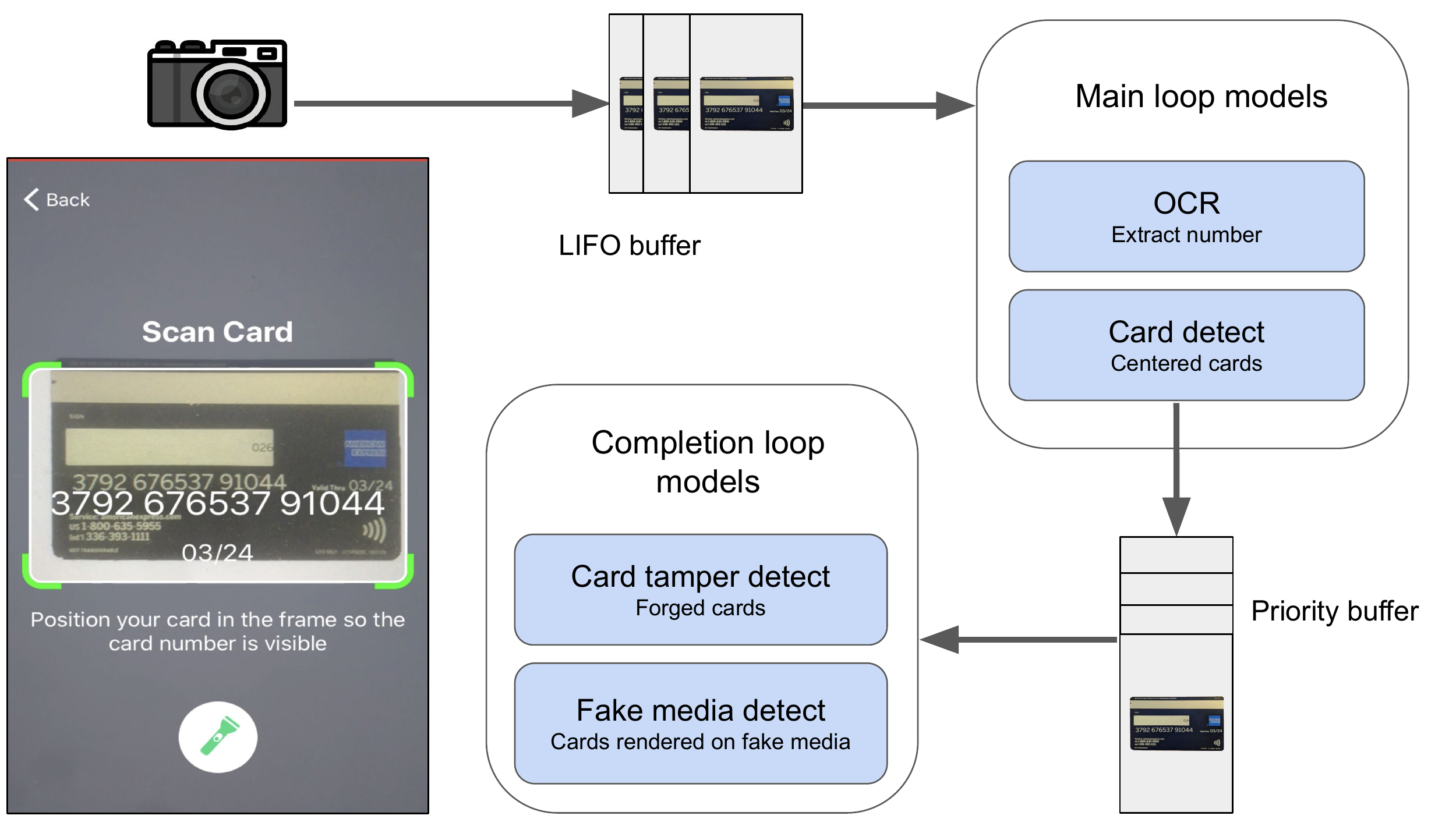}
\caption{Machine learning pipeline for client-side models.}
\label{fig:ml_pipeline}
\hrulefill
\end{figure}

Figure \ref{fig:ml_pipeline} shows our client-side machine learning
pipeline for processing images (frames) from the camera. This
pipeline uses two different producer/consumer modules and divides
the computation up into a \textit{main loop} and a \textit{completion loop}.
The main loop runs on images in real time as the camera extracts images, and
the completion loop runs after the main loop finishes but before making
the final API call.

In this flow we show the scanning process for a single side
of the card, but in \system we scan both sides of the card using the
same basic process before making the final API call. We introduce a
card detection model that detects the side of the card, which we use
as the basis for our two-side scan.
See Section~\ref{sec:design:redundancy_and_efficiency} for more details.

\section{Design}
\label{sec:design}
\subsection{Challenge: Where to run verification?}
Card verification can either run on the client or on the server.
Server-side verification moves compute intensive machine learning
inference away from the edge. This server-centric architecture ensures
verification can run on all phones, regardless of their compute
capabilities while also simplifying the role of the client to merely
relay data to the server. However, server-side verification puts
higher strain on network bandwidth and latency, with the need to transmit
frames from the camera to the server, resulting in delays in
verification.

Server-side verification also disregards end-user privacy. With
server-side verification, the app sends sensitive user information,
such as card images, to the server, thereby introducing potential
avenues for data breach.

Running verification on the mobile client involves running compute
intensive machine-learning inference on the client and only sending
high-level features to the server. This client-first architecture
puts less strain on the network and can process more frames faster by
virtue of running closer to the camera. Importantly, client-side 
verification is
more respectful of end-user privacy since it avoids sending sensitive
card images to the server.

\subsection{Solution: Run verification on the client} 
We believe that there are more good users than fraudsters and
respecting the good user's privacy should be the foremost concern for
anyone attempting to combat fraud. Additionally, one way fraudsters
source stolen card information is through data breaches, and we strive
to minimize these avenues. Thus, \system chooses to run its
verification on the client.  \system's system design and algorithmic
improvements ensure the running of uniform verification on resource-constrained
and well-provisioned devices across different platforms.

\subsection{Challenge: How to ensure high verification accuracy on a mobile phone?}

The input to our models is an image or a video stream of a user
holding a card.  Changes in illumination, varying camera quality,
orientation of the payment card, wear patterns on the card, and so on
add to the stochasticity of the inputs, which makes it difficult to
ensure high accuracy.  However, since we use this input to verify or
block a user, ensuring high accuracy is critical to provide uniform
verification.

A common solution to ensure high accuracy in machine
learning is to increase the model size. However, apps are hesitant to
increase the size of their binary \cite{play_store_apk_size},
mobile networks can be slow and content distribution networks are
expensive (a 5MB machine learning model downloaded 50 million times in 
a month costs north of \$30k / month) complicating model downloads in
the background. All of which puts pressure on client-side machine
learning to keep model sizes down while still providing fast and
accurate predictions.

\subsection{Solution: Decompose verification to sub-tasks for improved efficiency and redundancy}
\label{sec:design:redundancy_and_efficiency}

We decompose card verification into multiple tasks, with each task
having its own independent machine learning model. Decomposition of
the verification process into sub-tasks keeps each sub-task efficient
while also providing redundancy across tasks for improved accuracy.
Decomposition also enables us to iteratively refine models for each
individual task	until the models reach an acceptable level of accuracy.

\system decomposes verification into four distinct sub-tasks: OCR,
card detection, fake media detection, and card tampering detection. OCR
scans the number side of the card and extracts the card number, card
detection detects frames where the user centers the card in the viewport 
and detects the side of the card that the user scans (number or
non-number side), and fake media detection checks both sides of the
card to detect cards scanned off fake media such as device screens, paper, cardboard etc.

Card tampering detection also scans both sides of the card to detect signs
of tampering and inconsistencies. We scan both sides since newer card
designs have meaningful information printed on both sides.
For instance, newer Wells Fargo payment cards contain the bank and
payment network logos on one side and the card number and expiry 
on the other side. In this case, if the card tampering detection
detects a Wells Fargo card number on one side and detects a conflicting
bank logo on the same or opposite side, \system flags the scan as fake.

Decomposition leads to higher accuracy in two ways. First, our
decomposition makes our overall system more efficient, allocating
limited ML resources towards the images that are most likely to
generate meaningful signals (Section
\ref{sec:design:efficient}). Second, our decomposition provides
redundant signals to increase the confidence of the predictions that
\system makes (Section \ref{sec:design:redundancy}).

\subsubsection{Efficiency with decomposition}
\label{sec:design:efficient}
If we pass every frame coming from the camera through all our
machine learning models, then we waste computation. For example, if
there is an image without a card in it, then running the fake media
detection model or the card tampering detection model on that image is
wasteful because there isn't even a card in the image, and it won't
provide meaningful results.

Instead, to make our overall ML pipeline more efficient, we divide
computation up into a \emph{main loop} that runs on all frames in
real-time, and a \emph{completion loop} that defers running of models
and operates on only a subset of the frames that we believe are most
likely to have relevant fraud signals. Logic in the main loop dictates
which frames it passes on to the completion loop, which in \system are
any images that have centered cards in them. Figure
~\ref{fig:ml_pipeline} shows \system's decomposition.

At the heart of our design is the card detector model. The card
detector model is a 3-class image classifier that we train to detect
a centered image of the number side or a centered image of the
non-number side of a card. The card detector also has a third class,
called the background class, to filter out frames that contain
off-center cards or no cards at all.

We execute the card detector and OCR models on the main loop.
The reason that we run these models on the main loop is because they
both produce user-visible outputs (Figure \ref{fig:user_flow}). The
card detection model highlights the corner of our viewport when it
detects a centered card and our OCR model displays the recognized card
number and expiration date using an animation as it captures
them. Thus, these models must run in the main loop to process frames
in real-time and display their results to the user. We finish the main
loop by using the results from the card detection model to determine
when the user scans either the number side or non-number side of a
card for 1.5 seconds.

We execute the fake media detection and card tampering detection models on
the completion loop. These models only produce a result that our
system uses to detect fake cards via an API call, so we defer
execution until after the main loop finishes and only run them on a
subset of frames (up to six in our current system) identified by the
card detector model that are likely to produce evidence of fake cards.
Our decomposition keeps the system efficient by having the
completion loop save computation by only processing frames with
centered cards.

\subsubsection{Redundancy with decomposition}
\label{sec:design:redundancy}
\system uses different forms of redundancy for each of its models to
provide high confidence in the accuracy of its decisions. Some models
have a built-in validation signal for redundancy, while others require
external validation signals for redundancy.

More concretely, OCR has redundancy built into its design from
the Luhn algorithm~\cite{luhnCheck}.  The Luhn algorithm is a checksum used to validate
credit card numbers. Thus, we validate OCR predictions by making sure
that they satisfy the Luhn checksum.

In contrast, our card tampering detection model detects prominent
objects on cards (e.g., the Visa symbol) and our fake media detection
model detects cards scanned off fake media and do \emph{not} contain a
built-in validation signal. Thus, we use the predictions of the card
detection model and OCR to provide redundancy. Correlating predictions
between models reinforces their decisions. For example, predictions of
seeing a card by the card detection model, and detecting the presence of a Visa
symbol by the card tampering detection model reinforce each other. For the number
side, these predictions also reinforce OCR and in turn OCR reinforces them.

Additionally, OCR, card tampering detection, and fake media detection
benefit from voting on predictions across the frames they process
for redundancy. For example, if our fake media detection model processes five frames
and predicts the presence of a computer screen on three of them, and no screen on
the remaining two, its final decision is that a screen is present.

\newcolumntype{A}{>{\centering\arraybackslash}m{1.4cm}}
\newcolumntype{B}{>{\centering\arraybackslash}m{2.9cm}}
\newcolumntype{C}{>{\centering\arraybackslash}m{4.2cm}}

\begin{figure}[t]
  \centering
  \small
  \begin{tabular}{|A|B|B|}
    \hline
    {\bf Task} & {\bf Redundancy used} & {\bf Redundancy provided} \\ \hline \hline
    Card detection & None & Centered and focused card present \\ \hline
    OCR & Luhn + voting & Card number and location \\ \hline
    Card tampering & Voting + validation from card detection and OCR & None\\ \hline
    Fake media detection & Voting + validation from card detection & None \\ \hline
  \end{tabular}
  \caption{Task-level redundancy in \system.}
  \label{fig:decomposition_summary}
  \hrulefill
\end{figure}

Figure ~\ref{fig:decomposition_summary} summarizes the different forms
of redundancy we use with each model.

Redundancy is the most important lesson learned from our
implementation. Even if a model achieves an accuracy of 100\% on a
benchmark validation dataset, it can still fall short for a practical
system. Instead, one needs to supplement these predictions with
additional data via voting and validation signals. To cope with the
uncertainty inherent in real deployments and to handle active attackers,
we need these forms of redundancy.

\subsection{Challenge: How to account for resource-constrained mobile phones?}
Owing to differences in sensor quality and compute capabilities, there
is a stark difference in the performance of running image processing
machine learning tasks on resource-constrained and well-provisioned phones.  At best, the
result of the difference in this performance inconveniences users by
making them wait longer to verify their cards, and at worst, prevents
users from verifying themselves.  In either case, fraud systems penalize
users attempting to verify themselves simply for not possessing a 
well-provisioned phone.

From our measurement study (Section ~\ref{sec:measurement_study}), 
we can see first-hand the stark differences in running the same 
machine  learning models on well-provisioned and resource-constrained devices in a 
production setting.  Even though machine learning inference is 
expected to improve with streamlined accelerated hardware support 
(GPUs, Neural Engine) on iOS which will bridge the gap between
resource-constrained and well-provisioned iPhones, it continues to be a problem on 
Android phones due to inherent hardware heterogeneity, with over 
2000 SoCs in distribution, making optimizing for each of them 
difficult.

Thus, to have uniform verification on all devices
irrespective of hardware capabilities, there is a need
for software enhancements for efficient machine learning 
inference.

\subsection{Solution: Refine machine learning models and improve system design to provide faster effective frame rates}

Our solution to account for resource-constrained phones consists of algorithmic
machine learning improvements for faster inference times and refined
system design for higher utilization of the hardware.

\subsubsection{Improvements in machine learning}
\label{sec:design:ddocr}
The following two key principles inform our machine learning re-design:

\textbf{(1) Optimize machine learning for resource-constrained phones:} Machine
learning optimization for resource-constrained phones translates to well-provisioned phones
as well but the reverse is not true. Well provisioned phones often employ hardware
acceleration optimized for efficient machine learning inference. Having
this hardware support means that we can increase the capacity of machine learning models  
either by adding more parameters or by breaking a problem into sub-problems
each executed with a separate machine learning model. This has a sub-linear
slow-down in performance, leading to a better speed versus accuracy trade off. 
However, resource-constrained phones do not possess this luxury and adding
parameters to the model has at least a linear slow-down in performance
(it was quadratic in our case).

We thus create a unified model for OCR and reduce the number of
parameters by half.  This leads to a quadratic speed up on resource-constrained
phones and close to a linear speed up on well-provisioned phones as well. The
new model also occupies half the disk and memory space of the original model, 
as an added benefit to memory constrained devices.

In addition to the algorithmic improvements, using a single model avoids
expensive and complex processing to convert the output of one model
into the input of another, leading to a more efficient implementation
with less code needed to interpret the results.

\textbf{(2) Optimize machine learning for the common case:} Following
our previous design principle of using a single model for OCR implies
that we are operating at half the machine learning capacity as before
leading to an inevitable tradeoff between accuracy and speed.  We
observe that with a unified model for OCR we need to add complex
auxiliary layers at multiple stages in the model to scan all
payment card designs. However, these auxiliary layers add
parameters to the model as well as increase the post processing
complexity making them prohibitively slow on resource-constrained devices.

We thus add native support in the model for the most common designs
and employ system design strategies to account for less common card
designs. This ensures our machine learning inference is efficient for
the common case employing gated execution of more complex pipeline for
less common cases.

\textbf{OCR model design:} With the above two design principles, we
design and implement a new OCR model to work in a single pass. Our new 
model draws on ideas from existing work on Faster-RCNN~\cite{fasterrcnn},
SSD~\cite{SSD} and Yolo~\cite{yolo}.

We replace Boxer's detection and recognition stages, which were
implemented using two separate models, with a single network. The
network reasons globally about the entire image resulting in
end-to-end training and faster inference. We implement the model as a
fully convolutional MobileNetV2~\cite{mobilenetv2} with auxiliary
features for detection and recognition unifying separate components of
the detection and recognition into a single network. We append these 
features to the network at different layers to account for
multi-sized feature maps, like SSD~\cite{SSD}. This flexibility gives
us the ability to operate on credit cards with varied font sizes.

Our OCR model operates on an input image size of 600x375 pixels, which
is close to the aspect ratio of a credit card. As with any CNN,
the feature map shrinks in size and expands in the depth dimension as
the network processes the image. We add auxiliary layers to the
network at two places, one where the feature map size is 38x24, and
another where the feature map size is 19x12. We find adding
multi-layer predictions at these two layers captures the vast majority
of credit card fonts. The activations corresponding to feature map of
size 38x24 are useful for small and flat font payment cards while the
activations corresponding to the feature map size 19x12 are used for
embossed cards that have bigger fonts.

At the output feature maps, each activation is responsible for
detecting a digit. To extract the card number from an image, we need
to localize and recognize individual digits. Knowing the location and value 
of each digit in the input image aids in post processing to remove false
positives. Accordingly, each activation in the two output feature maps
is mapped to a regression layer (for localization) and a
classification layer (for recognition). We implement the regression layer 
with anchor boxes like Faster-RCNN~\cite{fasterrcnn},
where the possible output locations are captured with
multi-aspect-ratio bounding boxes. Unlike Faster-RCNN which uses nine
anchor boxes per location, we only use three, since we find this to be
sufficient for OCR. We also fine-tune our bounding box scales for
OCR; however, we defer these details to the open source code we make
available. To each output feature map activation, we append a
regression layer that consists of mapping each input activation to 12
output activations, since we output three bounding box proposals each
containing four coordinates.  Each of these proposals (bounding boxes)
can contain a digit that the classification layer detects. The
classification layer maps each input activation to 33 output
activations, 11 activations (background, 0 to 9) per bounding box.

During inference we apply standard post processing techniques like
non-max suppression~\cite{richfeature} and heuristic based refining
that is relevant to different credit card designs.

Our OCR model has difficulty in localizing small objects
precisely, much like Yolo~\cite{yolo} and SSD~\cite{SSD}. Since each
output activation is responsible for detecting a single digit, if the
corresponding receptive field of a single activation spans multiple
digits, the model will only be able to detect a single digit. In our
experience, we found one credit card design (Brex credit cards) that
the model struggles to perform OCR on. One way to fix this corner case
is to make the input feature map size bigger or add auxiliary layers 
earlier in the network where the feature map sizes are bigger. However, 
this adds more computation to the machine learning inference effectively 
decreasing the frame rates on resource-constrained devices.

To successfully perform OCR on payment cards with tiny fonts, 
we first detect the ratio of the size of an individual digit 
compared to the size of the input feature map. If it is below 
our empirically determined threshold, we pass a zoomed in image 
of the input through the machine learning pipeline effectively 
mapping a card with small font to one with a relatively bigger 
font that the model supports natively. This flow adds latency 
to our overall inference pipeline; however, \system only needs to 
trigger it sparingly.

We use 1 million real and synthetic card images to train our OCR model.
However, we find that, on our internal benchmark datasets, this model 
is unable  to  reproduce Boxer OCR's precision and recall, owing  to the 
overall reduced number of parameters. To account for this reduction, 
we generate an additional 1.5 million
synthetic credit card images. Ultimately, we
train the OCR model with 2.5 million real and synthetic card images
to match Boxer OCR's baseline on our benchmark datasets.

\subsubsection{Improvements in system design}
To further increase the frame rate, we refine our system design to use
a producer/consumer pipeline with a bounded buffer. We collect multiple
frames from the camera and run machine learning inference on all of
them in parallel. Since running machine learning inference takes time,
this design ensures that fetching frames from the camera is not
blocked, making the entire system parallel from reading camera frames
to completing machine learning inference.

%A common flow found in client-side machine learning inference,%especially on Android, is to capture a 
%frame from the camera and run the ML model while blocking on 
%the result, then repeating the entire process for a new frame. 
%The underlying assumption is that the camera frame rate is 
%fast enough that time spent waiting for a new
%frame is negligible and that machine learning inference is
%embarrassingly parallel, so one can extract most or all the
%performance through this blocking system design. 
We find that
buffering images and running the same inference in parallel leads to 
speedups of up to 117\% for our workload
(see Appendix \ref{appendix:prod-con} for our results).
Processing more frames is critical for improving the end-to-end success 
rate for complex machine learning problems that demand high accuracy 
as concluded in our measurement study (see Section \ref{sec:measurement_study}).

\newcommand{\daredeviltotalandroidapps}{70\xspace}
\newcommand{\daredeviltotaliosapps}{44\xspace}

\newcommand{\daredeviltotalandroid}{477,594\xspace}
\newcommand{\daredeviltotalandroidtype}{722\xspace}

\newcommand{\daredeviltotalios}{1,102,666\xspace}
\newcommand{\daredeviltotaliostype}{28\xspace}

\newcommand{\ddsamsungtype}{302\xspace}
\newcommand{\ddsamsungtotal}{328,600\xspace}

\newcommand{\ddhuaweitype}{111\xspace}
\newcommand{\ddhuaweitotal}{42,619\xspace}

\newcommand{\ddxiaomitype}{38\xspace}
\newcommand{\ddxiaomitotal}{6,876\xspace}

\newcommand{\ddlgtype}{78\xspace}
\newcommand{\ddlgtotal}{22,952\xspace}

\newcommand{\ddgoogletype}{17\xspace}
\newcommand{\ddgoogletotal}{31,699\xspace}

\newcommand{\ddmotorolatype}{58\xspace}
\newcommand{\ddmotorolatotal}{18,407\xspace}

\newcommand{\ddoneplustype}{29\xspace}
\newcommand{\ddoneplustotal}{8,751\xspace}

\newcommand{\ddtailtype}{89\xspace}
\newcommand{\ddtotaltail}{17,690\xspace}
\newcommand{\ddtailvendors}{28\xspace}

\newcommand{\ddfailandroid}{55,093\xspace}
\newcommand{\ddfailios}{119,826\xspace}
\newcommand{\ddfailtotal}{174,919\xspace}

\section{Evaluation}
\label{sec:evaluation}
In our evaluation, we answer the following questions:

\begin{itemize}
  \item Does \system bridge the gap between low- and high-end devices?
  \item Does \system prevent fraud in the wild while remaining ethical?
  \item What is \system's false positive rate when scanning real cards
    and running anti-fraud models?
  \item Does our use of redundancy improve overall accuracy?
  \item What is the impact of back-end networks and data augmentation 
        on overall success rates?
  %\item How does \system compare against other card scanners?
\end{itemize}

\subsection{Does \system bridge the gap between low- and high-end devices?}
\label{sec:does_dd_bridge_gap}
In this section, we measure \system's performance for its most complex
and carefully designed machine learning model: OCR. Although OCR is a
critical part of our fraud system (see Section \ref{sec:eval:fraud}
for real-world results of using \system to stop fraud), in this
experiment we use OCR to help people add credit and debit cards to an
app more effectively by scanning instead of typing in numbers.

\subsubsection{Measurement Platform}
To measure \system's performance, we perform a correlation study by
making it available to third-party app developers and measuring the
success rate for the users of their live production apps.  For \system
Android SDK, we present results from anonymous statistics sent by
\daredeviltotalandroidapps apps that deploy our library from December
2019 to late November 2020.  For \system iOS SDK, we present results
from anonymous statistics sent by \daredeviltotaliosapps apps that
deploy our library from late July 2020 to late November 2020. 

\subsubsection{Testbed}
\system Android SDK ran on a total of \daredeviltotalandroid Android devices spanning a total of
\daredeviltotalandroidtype Android device types. 
This included \ddsamsungtotal Samsung devices
spanning \ddsamsungtype Samsung device types,  \ddhuaweitotal Huawei devices spanning \ddhuaweitype
Huawei device types, \ddxiaomitotal Xiaomi devices spanning \ddxiaomitype
Xiaomi device types, \ddlgtotal LG devices spanning \ddlgtype
LG device types, \ddgoogletotal Google devices spanning \ddgoogletype
Google device types, \ddmotorolatotal Motorola devices spanning \ddmotorolatype
Motorola device types, \ddoneplustotal OnePlus devices spanning \ddoneplustype
OnePlus device types and tail of \ddtotaltail devices, spanning \ddtailtype device types
and \ddtailvendors vendors.
\system iOS SDK ran on a total of \daredeviltotalios iOS devices spanning a total of
\daredeviltotaliostype iOS device types. 

\subsubsection{Task}
As before, \system prompts users to scan their credit cards. The task, UI and the control flow
is identical to the measurement study (Section \ref{sec:measurement_study:task})

%When invoked, it starts the camera and prompts users to place their card in the center of the viewport. \system's OCR processes frames obtained
%from the camera and attempts to extract the card number and expiry from the card. Upon success, 
%it displays the card number and expiry to the user and sends the scan statistics
%to our server. In case, the OCR is unable to extract the number, the flow doesn't time-out. Instead, we let the user cancel the scan which provides us an additional user-level metric that can be used
%to guide a future iteration.

\system consists of a single-stage OCR where both detection and recognition happen
in a single pass. The input image from the camera is processed and sent to the
OCR model which outputs a string of digits.

\begin{figure}[t]
  \centering
  \small
  \begin{tabular}{|l|c|c|c|c|}
    \hline
    {\bf Model} & {\bf Size} & {\bf \# } & {\bf \# 2D} & {\bf \# Depth-wise}\\
    {\bf        } & {\bf      } & {\bf params. } & {\bf Conv(s)} & {\bf Conv(s)}\\ \hline \hline
    \system  & 1.65MB  & 861,242  & 39 & 25 \\ \hline
    Boxer  & 2.94MB & 1,528,919  & 30 & 18 \\ \hline
  \end{tabular}
  \caption{Comparison of model parameters and architecture of \system (44\% fewer parameters) and Boxer.
  Developers using architectures similar to these models for other 
  applications can expect to see similar frame rates.}
  \label{fig:daredevil_vs_boxer_model_params}
  \hrulefill
\end{figure}

\system uses a fully convolutional MobileNetV2~\cite{mobilenetv2} with auxiliary
features for detection and recognition for OCR and occupies 1.65MB on disk. The OCR model
processes an input image of size 600x375 and generates 51,300 output values
which are used to detect and localize the information for extraction.
It has a total of 861,242 parameters of which 830,362 are trainable.
\system uses 44\% fewer parameters than Boxer.
Figure \ref{fig:daredevil_vs_boxer_model_params} shows
a comparison of the model parameters between Boxer and
\system.

We use the same inference engines (CoreML for iOS and TFLite CPU for Android) for \system as 
Boxer, detailed  in our measurement study 
Section \ref{sec:measurement_study}. Like Boxer, we
quantize all our models using 16-bit floating point weights.

\subsubsection{Results- Key Performance Metrics}
As before, we use the same definitions for \textbf{frame rate} and
\textbf{success rate} for our
performance metrics as in Section \ref{sec:measurement_study}.

We show the impact that the new \system OCR model (Section~\ref{sec:design:ddocr}) 
has on the overall scanning success rate. Our informal goal with \system was to
improve the success rates on Android to match Boxer iOS. \system
uses algorithmic machine learning improvements, empirical
accuracy-preserving optimizations, high fidelity synthetic data,
and an improved system design to achieve this goal.

\begin{figure}[t]
  \includegraphics[width=\columnwidth]{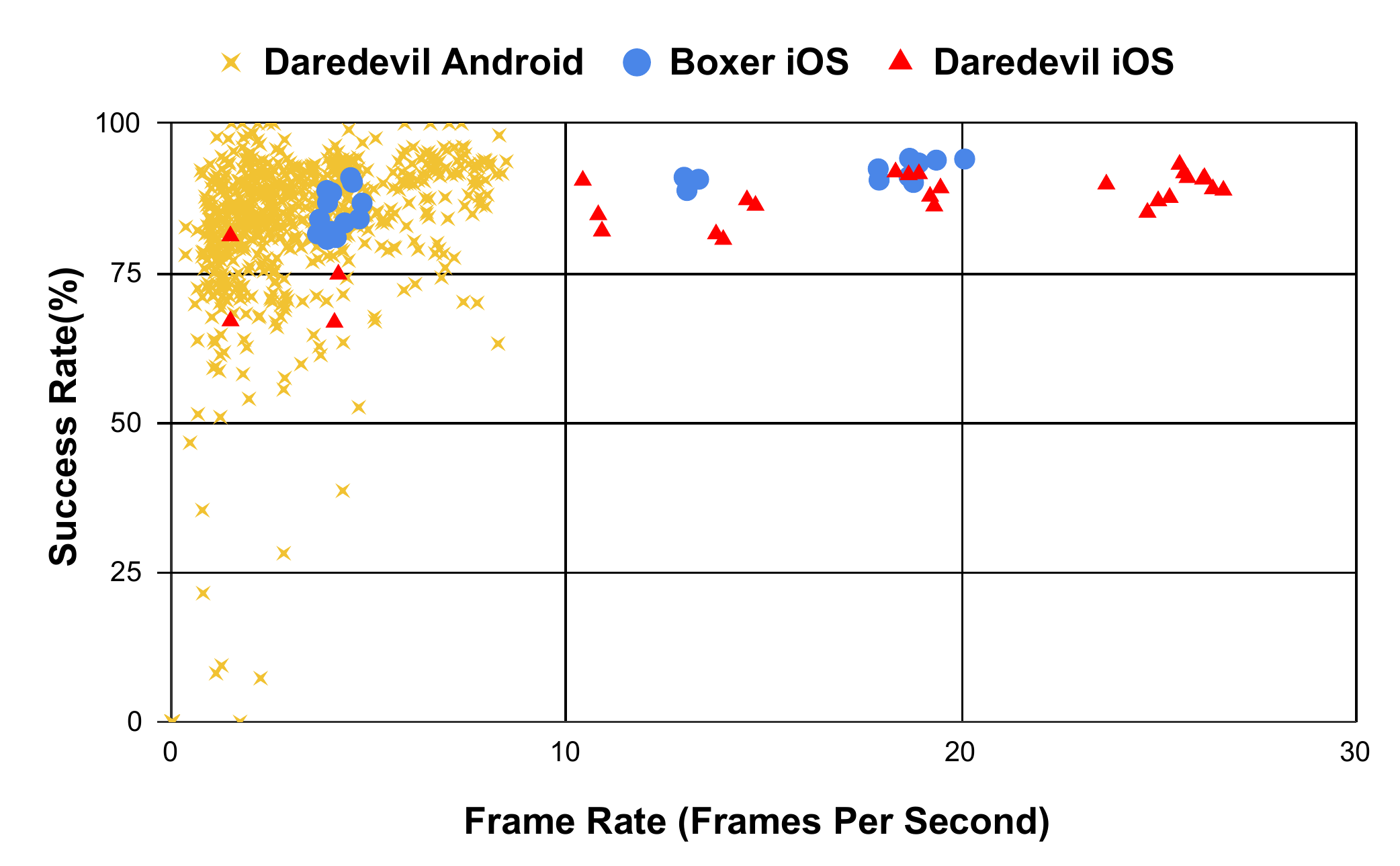}
  \caption{OCR success rate vs frame rate on \system Android, Boxer iOS
   and \system iOS.
    Each point is the average success 
    rate and frame rate for a specific device type. This figure 
    shows that by improving our machine learning model and 
    increasing the frame rate we can
    achieve higher success rates. The corresponding plot
    for Boxer Android is shown in Figure ~\ref{fig:boxer_ocr_visual}.}
  \label{fig:daredevil_ocr_stats}
  \hrulefill
\end{figure}

\begin{figure}[t]
  \centering
  \small
  \begin{tabular}{|l|c|c|c|c|}
    \hline
    {\bf Version} & {\bf Count} & {\bf Avg Suc} & {\bf Avg} & {\bf Avg}\\
    {\bf        } & {\bf      } & {\bf Rate} & {\bf FPS} & {\bf Dur (s)}\\ \hline \hline
    \system iOS  & \daredeviltotalios  & 89.13\%  & 20.00 & 9.37 \\ \hline
    Boxer iOS  & \boxertotalios  & 88.60\%  & 10.00 & 10.02 \\ \hline
    \system Android  & \daredeviltotalandroid & 88.46\%  & 4.07 & 10.55 \\ \hline
    Boxer Android  & \boxertotalandroid  & 46.72\%  & 1.30 & 15.45 \\ \hline
  \end{tabular}
  \caption{Comparison of \system and Boxer.
    %Card.io is an industry-standard card scanner that we use as a baseline. 
    We see, \system not only provides
    over 41\% improvement in success rates on Android but also improves iOS by close to 1\%.}
  \label{fig:overall_per_model}
  \hrulefill
\end{figure}

Figure \ref{fig:daredevil_ocr_stats} shows the results of \system
deployed on Android (which we refer to as \system Android) and 
iOS (which we refer to as \system iOS)
against Boxer iOS. This figure shows that \system's
improvements increase the success rate on Android to closely match
success rates on Boxer iOS, despite the massive hardware advantages present
on iOS. Seeing the success of \system Android, we ported it to iOS
and observed a more than 2x speedup in frame rates and a moderate 
improvement in the success rates as well (Figure~\ref{fig:overall_per_model}).
The increase in frame rates also lead to \system
being able to support iPhone 6 and below, which Boxer
does not support.

Concretely, from Figure~\ref{fig:overall_per_model} which
presents more detailed results, we see that the average frame rate
improves from 1.30 FPS on devices running Boxer Android to 4.07
FPS on devices running \system Android. \system Android also increases the 
average success rate from 46.72\% to
88.46\%. We also see an improvement in success rates on iOS,
going from 88.60\% on Boxer iOS to 89.13\% on \system iOS.
Additionally, the average scan duration decreases from 15.45s
to 10.55s on Android and from 10.02s to 9.37s on iOS. In our 
system we start the scan duration timer when the
user clicks on the ``scan card'' button and finish it after the scan
is complete, which includes accepting camera permissions, pulling
their card out of their wallet, scanning the card, and the 1.5s voting phase for 
error correction in the main loop.

\begin{figure*}[t]
  \centering
  \small
  \scalebox{1.0}{
  \begin{tabular}{|l|c|c|c|c|c|}
    \hline
    {} & \multicolumn{2}{|c|} {\bf \system} & \multicolumn{2}{|c|} {\bf Boxer} \\ \hline \hline
    {\bf Android FPS} & {\bf Count} & {\bf Success rate} & {\bf Count} & {\bf Success rate} \\ \hline \hline
    $<$ 1 FPS & 23,314 (4.88\%) & 37.92\% & 146,890 (44.61\%) & 31.87\% \\ \hline
    1$-$2 FPS & 48,271 (10.10\%) & 84.08\% & 97,798 (29.70\%) & 49.97\% \\ \hline
    $>=$ 2 FPS & 406,009 (85.01\%) & 91.88\% & 84,584 (25.68\%) & 68.72\% \\ \hline
  \end{tabular}}
  \caption{Success rates for Android devices running \system and Boxer by
    frame rate. We can see that \system significantly reduces the percentage of devices that
    operate below 1 FPS.}
  \label{fig:daredevil_android_frame_rate}
  \hrulefill
\end{figure*}

\system also improves the usability of card scanning with 4.88\%
(Figure \ref{fig:daredevil_android_frame_rate}) of Android phones
being able to process fewer than 1 FPS, compared to 44.61\% with Boxer
Android. Similar to Boxer,
the success rate for Android devices with less than 1 FPS (37.92\%) is
lower than the average success rate for Android overall (88.46\%),
however the overall increase in devices that can run the \system ML at
1 FPS or higher leads to a higher overall success rate (Figure
\ref{fig:overall_per_model} and Figure
\ref{fig:daredevil_android_frame_rate}). 

We see from Figure
\ref{fig:daredevil_android_frame_rate}
that for both Boxer and \system, as the frame rate increases the overall success
rate increases as well. Beyond 1 FPS, the success rate for \system witnesses
a precipitous rise compared to Boxer, this can be attributed to \system being 
trained with orders of magnitude more data (Section \ref{sec:eval:dataaug}), 
the use of an efficient machine learning pipeline (Section \ref{sec:design:redundancy_and_efficiency})
and marginal improvements seen from the updated back-end network (Section \ref{sec:eval:dataaug}).
It is clear that Boxer can also benefit from these improvements, however, given
that 44.61\% of the Android devices operate at below 1 FPS for Boxer (and \system also struggles
with devices that operate at frame rates below 1 FPS), a significant
portion of the devices will be excluded from these improvements. \system's architecture
reduces the number of devices that operate at below 1 FPS to 4.88\% which results in
significantly higher overall success rates.

%\textbf{Context for results.} For our fraud challenge we use OCR to
%verify the card number that the app has on record for this user. Thus,
%anyone who is unable to scan their number will be unable to pass the
%fraud challenge and increases in success rates lead directly to
%%increases in legitimate users who can verify themselves
%automatically. In Boxer's implementation, we saw that low-end 
%devices were unable to scan card numbers at a sufficiently high 
%frame rate despite the same machine learning models and the same system
%architecture. By improving our machine learning models 
%and system design with \system,
%we were able to scan cards on both low-end and high-end 
%devices with high success rates,
%regardless of differences in their hardware capabilities.

%\subsubsection{Further analysis of failure cases}
As with our measurement study Section ~\ref{sec:measurement_study}, we also
evaluated failed attempts with \system. We present the results in 
Appendix \ref{appendix:ddfail}.

\subsection{Does \system's fraud check work in the wild, while remaining ethical?}
\label{sec:eval:fraud}
\begin{figure}[t]
    \centering
	\includegraphics[width=0.9\columnwidth,height=5cm]{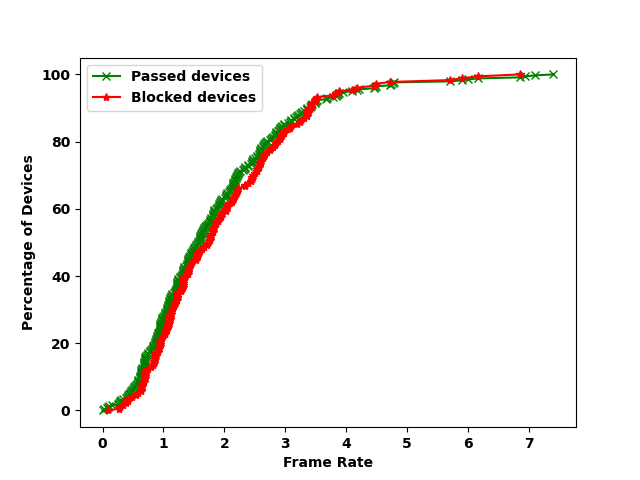}
    \caption{CDF of percentage of devices against the frame rates for devices passed 
    and blocked by \system. 
    We see that the two plots look very similar, indicating that \system's fraud decision is largely independent
    of the frame rate.}
    \label{fig:offerup_android_cdf}
    \hrulefill
\end{figure}

To evaluate \system's ability to stop fraud in real-time, we report
results from a large international app deploying our SDK. For a test
period of 3 months, the app flagged 12,474 transactions as suspicious
and challenged them with \system to verify their payment method.

\system passed 7,612 transactions and blocked the remaining 4,862
transactions.  Of the 7,612 transactions passed by \system, only 12
resulted in chargebacks, leading to a false negative rate of
0.16\%. We are unable to report the false positive rate since the app
did not share the false positive data with us, please see Section
\ref{sec:eval:false_positive} for a evaluation of the \system's false
positive rate.  Based on this initial test, the app has decided to
deploy \system.

To determine if \system's fraud decisions are correlated with the
device frame rates we further analyze the performance characteristics
of the passed and blocked devices.  We find that the average frame
rate of devices that \system passed was 1.84 FPS and the average frame
rate of the devices that \system blocked was 1.94 FPS, indicating that
the frame rates for the two groups is roughly the same. To visualize
these results, we plot the CDF of percentage of devices vs frame rate
(FPS) for the two groups and present the results in Figure
\ref{fig:offerup_android_cdf}. We see that the plots look very similar
indicating that frame rate is not a discriminating factor between the
blocked and passed groups.

For companies, chargebacks are the ground truth because they represent
exactly what they are liable for financially. However, it is possible
that there was fraud that happened but the victim failed to report the
fraudulent charge to their issuing bank, thus the actual amount of
fraud may be higher than the chargeback count that we report in this
experiment.

\subsection{What is \system's false positive rate when scanning real cards
    and running anti-fraud models?}
\label{sec:eval:false_positive}

To evaluate \system's false positive rate, we report results from
four authors scanning 105 cards in a lab setting using the latest
production anti-fraud models as of December 2020. In this experiment,
we invoke the fraud flow and record the number of scans that the
system incorrectly flags as being fraudulent. This section
complements our real-world evaluation of \system's fraud systems in
~\ref{sec:eval:fraud} that shows our false negative rate.

We scan 105 different real cards multiple times on different 
resource-constrained and well-provisioned Android and iOS devices 
for a total of 310 scans. The devices we use are 
iPhone SE (1st gen), Google Pixel 2, Nexus 6, iPhone 6s, and 
iPhone 11. Of these 310 scans, \system incorrectly flags seven 
scans as fraudulent, giving a false positive rate of 2.2\%. 
The false positives are uniformly spread across all devices,
indicating that \system does not unfairly permit well-provisioned or 
resource-constrained devices, similar to our fraud decisions as discussed 
in Figure~\ref{fig:offerup_android_cdf}.

Six out of the seven reported false positives were transient in nature, i.e.
further scans of the same card (which we would expect from a good user)
did not result in false positives. The other card was consistently
flagged incorrectly by our fake media detection model.

\subsection{Does our use of redundancy improve overall accuracy?}
\label{sec:eval:user_study}
In this section, we evaluate the effectiveness of our redundancy based
decomposition strategy (described in Section~\ref{sec:design:redundancy_and_efficiency}) 
in aiding fraud detection. Specifically, we  evaluate the gains in accuracy on executing 
our card tampering detection and fake media detection models in the completion loop.

We run a user study with and without the card
detection model in the main loop to show how it benefits the card tampering
detection and fake media detection models running in the completion loop. The
user-facing feedback from the card detection model ensures that users
center their credit cards so that both models necessarily make their predictions
on valid credit card images.

Users participating in our study randomly run one of two versions of our app
and scan 30 different predetermined credit card images on a browser that we
provide via a link. We use their scans to evaluate the impact of the feedback
from card detection in terms of the number of mistakes made by card tampering
detection (i.e. objects present on the card that the model fails to detect as well
as objects not present on the card that the model incorrectly detects) and the accuracy 
of fake media detection in detecting both, the presence and absence of screens.

Our university's IRB board reviewed our user study and ruled it to be
exempt from IRB.

\begin{figure}[t]
  \centering
  \small
  \begin{tabular}{|l|c|c|}
    \hline
    {\bf    } & {\bf Card tampering} & {\bf Fake media}\\
    {\bf    } & {\bf detect. \# errors} & {\bf detect. acc.}\\ \hline \hline
    {\bf No Card Detection} & 1.94 errors per frame & 86.24\% \\ \hline
    {\bf With Card Detection} & 1.26 errors per frame & 95.26\% \\ \hline
  \end{tabular}
  \caption{Results from our user study indicate fewer errors made by the card tampering 
  detection model and higher accuracy of fake media detection model when aided by
  card detection.}
  \label{fig:redundancy_impact_summary}
  \hrulefill
\end{figure}

Figure~\ref{fig:redundancy_impact_summary} summarizes our results. Our design of decomposition centered on the card detection model ensures that we pass high-quality frames to the machine learning models, resulting in fewer errors for the card tampering detection model, decreasing the errors per frame from 1.94 errors per frame down to 1.26 errors per frame. This change also improves accuracy for our fake media detection model increasing the accuracy from 86.24\% to 95.26\%. Overall, these improvement lead to more accurate fraud detection. For more details on our user study, please see Appendix \ref{appendix:user_study}.

\subsection{What is the impact of back-end networks and data augmentation 
            on overall success rates?}
\label{sec:eval:dataaug}

\begin{figure}[t]
  \centering
  \small
  \begin{tabular}{|l|c|c|c|c|c|}
    \hline
    {\bf Back} & {\bf Size} & {\bf No. of } & {\bf Recall} & {\bf Precision} & {\bf FPS on}\\
    {\bf End    } & {\bf     } & {\bf Params. } & {\bf} & {\bf} & {\bf Pixel 3a}\\ \hline \hline
    MBv1  & 1.8MB  & 869,754  & 54.06\% & 100\% & 7.19 \\ \hline
    MBv2  & 1.65MB & 861,242  & 56.25\% & 100\% & 7.09\\ \hline
  \end{tabular}
  \caption{Comparison of model parameters and accuracy metrics on our benchmark 
  datasets using Daredevil with back-ends MobileNet V1 (MBv1) and MobileNet V2 (MBv2).
  We can see that using MobileNet V1 as back-end leads to less than 1\% 
  increase in model parameters with no decrease in precision and marginal decrease in recall.
  It should be noted that the second model (with back-end MobileNet V2) is currently in 
  production, all the statistics from \system evaluation correspond 
  to this model.}
  \label{fig:v1vsv2}
  \hrulefill
\end{figure}

To quantify the impact of back-end networks, we validate our models 
on image frames extracted from videos recorded by users scanning 
their credit cards. Crucially, this is the same benchmark we use to
evaluate models that are shipped in production. The test set consists 
of 640 image frames extracted from 32 videos. We train \system OCR 
with MobileNet V1 and MobileNet V2 back-ends and report the results in 
Figure \ref{fig:v1vsv2}. We define a correct prediction as one where the 
model can correctly extract the card number from the image frame, while 
an incorrect prediction is one where the model extracts an incorrect card 
number (valid but incorrect), finally all frames where the model is able to 
extract only a partial number are considered missed predictions. Accordingly, 
recall is the fraction of the frames where the model produced a
correct prediction and precision is the fraction of the all the predictions
that were correct.

\begin{figure}[t]
  \centering
  \small
  \begin{tabular}{|l|c|c|}
    \hline
    {\bf No. of images } & {\bf Recall} & {\bf Precision}\\ \hline \hline
    %{\bf } & {\bf} & {\bf}\\ \hline \hline
    495,134  & 20\% & 98.46\% \\ \hline
    939,165  & 27.96\% & 98.89\% \\ \hline
    1,374,707  & 42.08\% & 99.26\% \\ \hline
    2,006,452  & 49.06\% & 100\% \\ \hline
    2,500,612  & 56.25\% & 100\% \\ \hline
  \end{tabular}
  \caption{Impact of varying the amount of training data on model accuracy.
  The model consists of \system OCR with MobileNet V2 back-end.}
  \label{fig:impact_data_aug}
  \hrulefill
\end{figure}

Critically, from Figure \ref{fig:v1vsv2} we see that using MobileNet V2 instead of MobileNet V1 as the back-end
network results in less than 1\% reduction in the number of parameters, indicating 
that the reduction in the overall parameters is a direct result 
of \system's architecture independent of the back-end network. We also see \system
with MobileNet V1 closely matches \system with MobileNet V2 in recall and precision
(Figure \ref{fig:v1vsv2}) further highlighting the back-end agnostic nature of \system.

To quantify the impact of data augmentation on the improvement of overall success rates.
We train \system OCR (Mobile Net V2 back-end) by varying the amount of training data. Our 
training data is generated using a custom Generative Adversarial Network (GAN)
\cite{goodfellow2014generative} 
architecture and we also use standard data augmentation techniques in addition 
to the GAN. We evaluate the models using the same benchmark as before and report the results
in Figure \ref{fig:impact_data_aug}.

In summary, we conclude that with \system's architecture we are able to achieve the 
desired frame rate and with high-fidelity synthetic data we are able to achieve
the desired accuracy.

\section{Related work}
Our work is related to papers in the areas of financial fraud, challenge 
based authentication, computer vision, machine learning systems 
and machine learning for mobile.

Recent work has focused on devising challenges that rely on having 
users interact with their mobile phones to collect signals that are 
then processed for verification ~\cite{cardiocam, rtcaptcha}. 
Liu et al.  propose CardioCam ~\cite{cardiocam} to verify users based on 
their cardiac biometrics.  Researchers have also
devised authentication systems where users are challenged  to respond to
a Captcha challenge on their mobile phones, while collecting audio and
visual data of the response that is transmitted to a secure server for
processing ~\cite{rtcaptcha}. 

The execution of machine learning models on resource constrained
platforms such as mobile phones has seen active research in both
algorithmic machine learning improvements ~\cite{mobilenet, peleenet}
as well as enhanced system design ~\cite{facebook_ml, mcdnn,
  adadeep}. Liu et al. devise a selection framework, AdaDeep, that
automatically selects a combination of compression techniques to be
applied to a given neural network to balance between performance and
availability of resources ~\cite{adadeep}.  Closer to our work,
researchers at Facebook extensively profile the wide diversity in
compute capabilities on mobile phones for machine learning
~\cite{facebook_ml}. They also identify the benefits of optimizing to
run inference on CPUs over GPUs to provide stable execution on Android
devices, and \system follows this general plan where we run Android
models on the CPU but use the hardware acceleration available on iOS
to speed up our models.  Ran et al. \cite{deepdecision} create a
client-server hybrid framework to provide sufficient compute power for
running augmented reality apps. Authors in \cite{ogden2019characterizing}, 
\cite{deeplearningapps} conduct 
a measurement study of mobile performance analysis of various deep learning models and
conclude the need for extensive optimization and both on-device and cloud based inference.

Recently there has been work on improving the performance of parallel DNN 
training ~\cite{pipedream, gpipe}. Narayanan et el. ~\cite{pipedream} cast 
DNN training as a computational pipeline to efficiently utilize the hardware 
resources. In contrast, Huang et al.~\cite{gpipe}, while also using pipelining to 
train large models, significantly reduce the memory overhead by 
re-materialization. 

Apps such as Google Pay and Apple Pay are restrictive in the users they
allow to use their systems. Firstly, they are not available in all regions
around the world ~\cite{google_pay_availability}, ~\cite{apple_pay_availability}.
More importantly, these services are restrictive in their support to pre-paid
cards ~\cite{google_pay_prepaid}. Over 8 million households in the United States rely on pre-paid cards, 
most of whom are blocked from using these services ~\cite{low_ses_bills}.

Payment card fraud using card skimmers has been studied recently by 
Scaife et al.~\cite{scaife}. In this work, researchers built a card skimmer 
detector that can be used at physical payment terminals such as ATMs and gas 
stations. In another work, Scaife et al.~\cite{kiss} did a survey of gas 
pump card skimmer detection techniques including Bluetooth skimmer detection 
on iOS and Android apps, to identify common skimmer detection characteristics.

\section{Conclusions}
Deep learning has seen a widespread adoption in a multitude of domains, 
outperforming traditional machine learning and rule-based algorithms. We 
have also seen it make in-roads into security with its potential to 
empower data engineers with newer features that can limit the prejudices
of prior algorithms. However, if not careful, deep-learning-based
security challenges have the potential of reproducing historical prejudices, 
improving the security and user experience of one group at the expense 
of altogether blocking the other.

In this paper, with a wide-scale measurement study consisting of
\boxertotal devices that ran in real apps. Our study looked at a
widely deployed deep-learning-based system for scanning payment cards
where we demonstrated that while these challenges can solve the app's
business problem by functioning reliably on high-end phones, this
challenge has the potential to disproportionately block users from low
socio-economic tiers who rely on lower tier smartphones.

With the lessons learned from our measurement study, we designed \system, 
a payment card verification system that used deep learning 
optimizations and improved system design to build a complex 
security system that works uniformly on low-end and high-end 
mobile devices. We showed the results from
\daredeviltotal devices from \system's public deployment to demonstrate
the practical nature of our system across all devices.

\section*{Acknowledgments} 
We would like to thank Xiaojing Liao and the anonymous reviewers 
who provided valuable feedback on this work. We would also like to the 
thank Weisu Yin, Sven Kuhne and Allison Tearjen for their contributions to this work. 
This research was funded by a grant from Bouncer Technologies.
%\begin{center}
%  {\large\bf Appendix\vspace{-.5em}}%
%\end{center}
\appendix

\subsection{How does \system compare against other card scanners?}
\label{sec:other_card_scanners}
Card.io~\cite{cardio} is a popular open-source scanning library commonly used
in the industry. We compare \system against Card.io via a lab experiment to
measure their scan success rates on our benchmark test set of 100 credit cards.
We observe that \system is able to extract the correct
card number from each card, while Card.io is able to extract the correct
card number from only 58 cards. Accordingly, \system's precision and recall
are both at 100\%, while Card.io's precision and recall are 100\% and 58\% respectively.
The lower recall of Card.io is attributed to its inability to scan cards with flat fonts.

\subsection{Impact of the producer / consumer design on frame rates.}
\label{appendix:prod-con}

\begin{figure}[t]
  \centering
  \small
  \begin{tabular}{|l|c|c|c|}
    \hline
        {\bf Device} & {\bf Blocking} & {\bf + Buffer} & {\bf + Parallel} \\ \hline \hline
        iPhone 5s & 1.65 fps & 1.70 fps & 2.95 fps \\ \hline
        iPhone SE & 7.60 fps & 7.90 fps & 14.90 fps \\ \hline
        iPhone XR & 28.45 fps &  32.60 fps & 32.60 fps \\ \hline
        LG K20 Plus & 1.03 fps & 1.04 fps & 1.39 fps \\ \hline
        Xiaomi Redmi 7 & 3.16 fps & 3.47 fps & 4.89 fps \\ \hline
        Pixel 2 & 3.66 fps & 4.35 fps & 7.95 fps \\ \hline
  \end{tabular}
  \caption{Frames per second for 20 second run. This figure shows the
    performance improvement measured by frames processed by our main
    loop per second with the baseline of a blocking system, a system
    that buffers images, and a system that buffers images and runs the
    ML models in parallel.}
  \label{fig:prod_cons_overall_results}
  \hrulefill
\end{figure}

%\begin{figure}[tb]
%    \centering
%    \begin{subfigure}[t]{0.485\columnwidth}
%        \centering \includegraphics[width=1\textwidth,
%          height=1\textwidth]{figs/idle_pipeline.pdf}
%        \caption{Figure showing the idle time slots while the CPU and
%          GPU wait for each other to finish.}
%        \label{fig:idle_pipeline}
%    \end{subfigure}\hfill%
%   ~
%    \begin{subfigure}[t]{0.485\columnwidth}
%        \centering \includegraphics[width=1\textwidth,
%          height=1\textwidth]{figs/Blocking.png}
%        \caption{Plot showing the GPU utilization for an interval of
%          1.8 seconds with blocking OCR.}
%        \label{fig:blocking_ocr_gpu}
%    \end{subfigure}
%    \caption{}
%    \label{}
%\end{figure}
%
%\begin{figure}[tb]
%    \centering
%    \begin{subfigure}[t]{0.485\columnwidth}
%        \centering \includegraphics[width=1\textwidth,
%          height=1\textwidth]{figs/full_pipeline.pdf}
%        \caption{Figure showing significant reduction in idle time
%          slots with the use of producer consumer style design.}
%        \label{fig:full_pipeline}
%    \end{subfigure}\hfill%
%   ~
%    \begin{subfigure}[t]{0.485\columnwidth}
%        \centering \includegraphics[width=1\textwidth,
%          height=1\textwidth]{figs/NonBlocking.png}
%        \caption{Plot showing the GPU utilization for an interval of
%          1.8 seconds with non-blocking producer consumer style OCR.}
%        \label{fig:non_blocking_ocr_gpu}
%    \end{subfigure}
%    \caption{}
%    \label{}
%\end{figure}

\system processes frames obtained from a live camera feed. In most
cases, the camera runs at a higher frame rate than the machine
learning model, meaning that applications will have to drop some
number of frames while the user is scanning their card. A natural and
common solution to this problem is to block the live feed while the
prediction runs, waiting for the machine learning models to finish
processing the frame before grabbing the next available frame from the
camera. This solution leads to a lower effective frame rate because of
the waiting time but is memory efficient and ensures that the models
always have fresh data to process by virtue of using only the latest
frame from the camera.

As opposed to processing each camera frame serially and blocking
the live feed while the model runs, \system uses a 
producer (the camera) / consumer (machine learning models) 
architecture with a bounded LIFO buffer to
store the most recent frames, and run multiple predictions in
parallel. This architecture comes at the increased cost of memory but
enables the machine learning models to execute without any waiting and
ensures that the models process frames that are close to what the user
sees.

We have already seen the producer/consumer design leading to higher frame rates
and success rates in production (Section ~\ref{sec:does_dd_bridge_gap}).
In this section, we run a controlled lab experiment to compare the
frame rates between the blocking design and producer/consumer design
of running our main loop (the card detection and OCR models) on frames
produced from a fixed camera feed. We follow this up with a qualitative
analysis of why the blocking design is slower on both Android and iOS,
despite the considerable differences in how the two platforms execute
machine learning inference.

Specifically, we consider three different variations: (1)
a blocking style with a single instance of our main loop models driven
at the frame rate of the camera, (2) a non-blocking style using a buffer
to store the two most recent frames with a single thread running our main
loop models, and (3) a non-blocking style using a buffer to store the two
most recent frames with two threads on iOS and four threads on Android
running our main loop models. We run this experiment by measuring the
frame rates observed on running the three variations on different iOS
and Android devices of varying capabilities for 20 seconds each.

Figure ~\ref{fig:prod_cons_overall_results} summarizes our results from these
experiments. From these results we can see a clear increase in the frame rates
across all phones on both iOS and Android on moving from a blocking
system to a system that buffers frames to a system that buffers frames
and runs our main loop models in parallel.

Improvements in frame rates due to buffering alone range from 1\% to
19\%, with faster devices seeing larger gains. The reason that faster
devices see larger gains is because the time spent waiting for a
camera frame is a larger portion of the overall execution time as the
time spent on machine learning predictions goes down.

Surprisingly, we observe speed ups ranging from 15\% to 117\% due to
adding multiple instances of our main loop models that run predictions
in parallel. This speed up is surprising because machine learning
inference is embarrassingly parallel and the underlying hardware
architectures for iOS and Android are vastly different, so we did not
expect to see gains in performance on both platforms from the same
architectural improvements.

On iOS as opposed to Android, our machine learning models run on the
GPU, however, the CPU needs to encode the work on a command buffer
before GPU can execute it. Blocking the live feed while the prediction
is running can lead to idle time, since the GPU has to wait for the
CPU to encode the task. 
%Conceptually, this looks like
%Figure~\ref{fig:idle_pipeline} and the corresponding GPU utilization
%is shown in Figure~\ref{fig:blocking_ocr_gpu}.

Our producer/consumer style OCR addresses the GPU idling issue by
creating parallel workloads which ensures that the CPU will encode the
next workload while the GPU is executing the current workload. The
producer pushes the frames from the camera feed onto a buffer keeping
the most recent frames and removing old stale frames. The consumer
which consists of independent machine learning analyzers pull images
from the buffer and run predictions on the frames in
parallel. Internally, Core ML (Apple's machine learning framework)
serializes the requests, however, with this style, encoding and
execution happens in parallel. 
%Conceptually, this looks like
%Figure~\ref{fig:full_pipeline} and the corresponding GPU utilization
%is shown in Figure~\ref{fig:non_blocking_ocr_gpu}.

\begin{figure}[t]
  \centering
  \includegraphics[width=1\columnwidth]{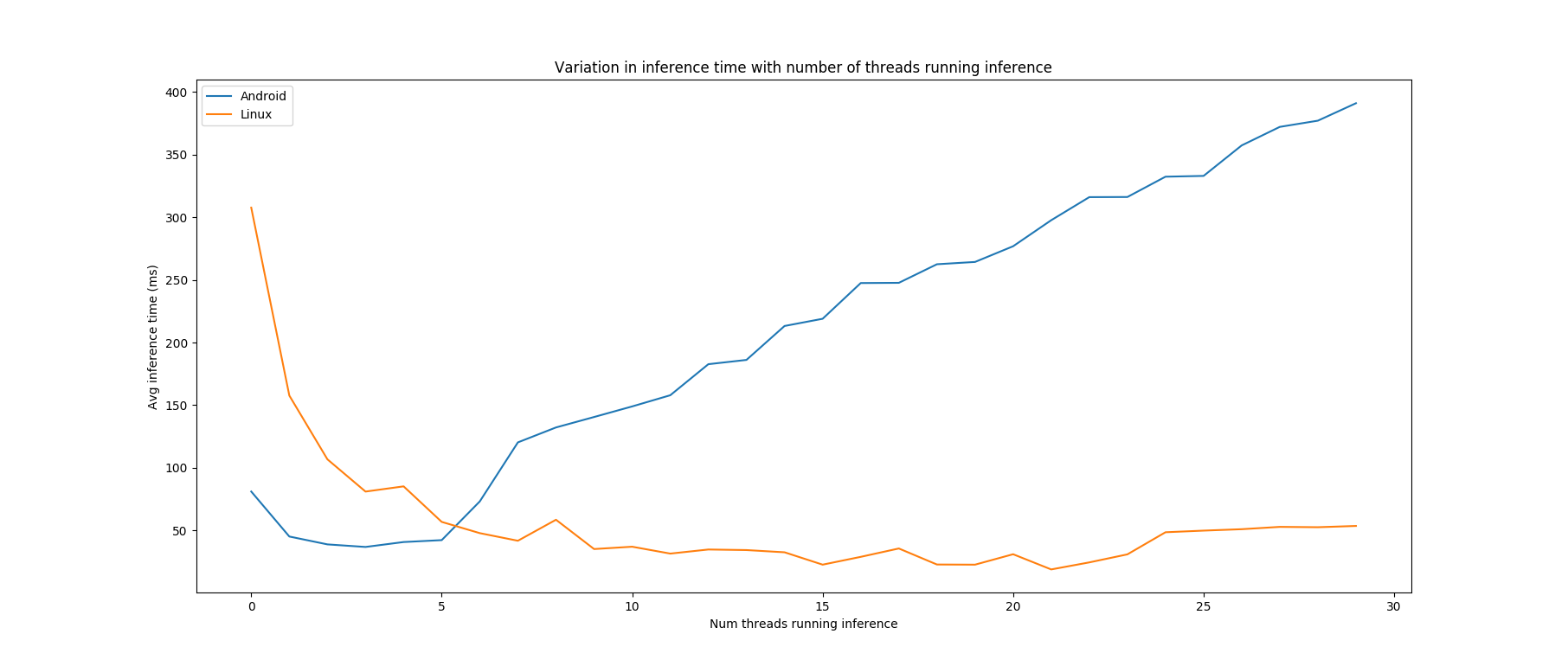}
  \caption{Variation in inference time with increasing number of
    threads for inference on a single TensorFlow lite interpreter
    on an 8 core Android device(big.LITTLE ARM) and a 24 core Linux server(x86).
    We attribute the increase in inference time on Android after 4 threads
    to the increased computation running on its slower cores. In contrast,
    the Linux server shows a continual decrease in inference time on
    running inference upto 24 threads, when all of its uniform cores
    are maximally utilized.}
  \label{fig:tflite_inf_time}
  \hrulefill
\end{figure}

On Android, since we run our machine learning models on the CPU, bubbles arising
as a result of the communication between the CPU and GPU are not
applicable to Android. Our producer/consumer OCR as well as a
sequential blocking OCR, both run multi-threaded machine learning
inference (using the industry standard TensorFlow Lite).  The two
differ in terms of the number of TensorFlow Lite interpreters running
inference, with the former using multiple independent interpreters and
the latter using a single interpreter. In this section, we seek to
understand why we observe higher frame rates with the producer/consumer
OCR. More concretely, we seek to understand the differences in how much
parallelism is available and how the hardware is utilized in both
cases to explain the improvements.

TensorFlow Lite runs machine learning inference by traversing a
computational graph where the nodes represent computations that are
part of the model and edges represent the dependence of values between
different computations \cite{tensorflow}. We inspect the TensorFlow
Lite source code to find that while running multi-threaded inference
through a single interpreter, it is only the individual computations
corresponding to the nodes of the computational graph that execute on
multiple threads, while the invocation of these nodes happens
sequentially on a single thread. Thus, with a single interpreter
not more than one node of the computational graph can run at a given time.

As a result, increasing the number of threads for a single interpreter
does not lead to faster inference if some threads execute slower than
others. While most Android phones in use today either have 4 or 8 CPU
cores, we uniformly see the optimal performance when using 4 threads for
inference. We attribute this to the adoption of Arm's big.LITTLE architecture
\cite{big-litte-arm} on phones with 8 cores, where 4 cores are designed for
efficiency rather than performance, and are thus slower than the other 4
cores designed for performance, while all cores are uniform in quad core
Android devices.

We verify the slowing down of inference on heterogeneous cores by plotting
the variation in inference time against the number of threads used on an
Android device having 8 cores following ARM's big.LITTLE architecture and
a Linux server having 24 equivalent cores running x86.  The inference times
start going up beyond 4 inference threads on the Android device, while it
starts to go up only after 24 threads on the Linux server.  These plots are
shown in Figure ~\ref{fig:tflite_inf_time}. 

Our producer/consumer OCR is not affected by the heterogeneity of CPU
cores since it invokes multiple interpreters in parallel. Multiple
nodes belonging to separate graphs coming from distinct interpreters
can execute at the same time, showing a better utilization of the
available hardware and correspondingly faster scan times with the
producer/consumer OCR.

\begin{figure}[t]
  \centering
  \small
  \begin{tabular}{|l|c|c|c|}
    \hline
    {\bf Platform} & {\bf 
    Count} & {\bf Avg } & {\bf Avg}\\
    {\bf        } & {\bf      } & {\bf FPS } & {\bf Duration (s)}\\ \hline \hline
    Daredevil Android  & \ddfailandroid  & 3.04  & 22.58 \\ \hline
    Daredevil iOS  & \ddfailios & 18.77  & 17.38  \\ \hline
  \end{tabular}
  \caption{Failure cases for \system on Android and iOS}
  \label{fig:dd_failure_android_and_ios}
  \hrulefill
\end{figure}

\subsection{Analysis of \system's failure cases}
\label{appendix:ddfail}
From our real-world deployment of \system, we observed \ddfailtotal
failed attempts where users gave up on trying to scan their card. We
aggregate the duration of these scans on iOS and Android to report
that Android users waited an average of 22.58s and iOS users waited an
average of 17.38s to scan their cards before giving up. This is shown
in Figure \ref{fig:dd_failure_android_and_ios}.

\subsection{User study to evaluate the use of redundancy based decomposition}
\label{appendix:user_study}
We ask users participating in our study to visit a link where they can scan
30 different credit card images via our app running on their phone. On opening
the link, we display 30 credit card images in a random sequence from a predetermined
set of cards. We manually label the objects present on these cards, such as bank
logo etc., for each frame collected from our user study videos which serve as ground
truth labels for card tampering detection. Although users participating in the
study scan cards that are displayed on device screens, we manually label each frame
from each video for the presence of screens to cover cases where the user starts
executing the app on the phone before pointing it at the screen. These labels are
our ground truth labels for fake media detection. The users randomly run one of two
versions of the app: with and without the card detection model. We carry out our user
study virtually due to the restrictions imposed by the COVID-19 pandemic.

We obtain a total of 603 scan videos from the user study, of which 273 were collected
by providing explicit feedback to the user to center their card by running the card
detection model and the remaining 330 were collected without any such feedback. For
the scans collected without feedback, we pass all extracted frames through the card
tampering detection and fake media detection models. For the scans with feedback, we
first pass the frames through the card detection model to only select those with centered
cards to pass to the card tampering detection and fake media detection models. We then
compare the performance of the two models in both cases.

From scans without feedback from card detection, we randomly sample 50 scans and
pass all 4,213 frames extracted	from them to the card tampering	detection model.
We consider expected objects not detected by the model as well as objects
incorrectly predicted by the model that are not present in the card as errors.
The model makes	a total	of 8,163 errors	at an average of 1.94 errors per frame.
We also	sample 50 random scans from those with feedback and pass 1,973 centered	
frames extracted from them to the card tampering detection model. In this case,	it
makes a	total of 1.26 errors per frame.

The fake media detection model makes correct predictions on 21,413 out of
24,829 frames extracted from all 330 scans without feedback at an accuracy
of 86.24\%. Of the 9,512 centered frames extracted from all 273 scans with
feedback, the fake media detection model makes correct predictions on 9,061
frames at an accuracy of 95.26\%. 

\subsection{Will increasing the frame rate further continue to increase the success rate?}
\label{sec:eval:framerate}
This section serves to answer the question of whether increasing our
current frame rates would lead to further improvements in success rate without changing the
machine learning model. Since card scanning involves sending frames
from a live camera feed through a machine learning model, faster frame rates
could imply two consecutive frames being practically identical to the eyes of a
machine learning model, leading to no gains obtained from a higher frame rate.
Alternatively, it could be that there are sufficient differences between two
consecutive frames for the machine learning model to produce a different and possibly
better prediction, resulting in a shorter scanning duration.

Concretely, consider an example where an OCR model is able to process
frames from the user's video feed at a rate of 5 FPS, and the user scans
for 10 seconds. This means that we run OCR inference on 50 frames in total.
We refer to the number of frames on which the model makes correct predictions
as the number of \emph{useful frames}. If this model makes correct predictions
on 10 frames, then we have 10 useful frames from the total set of 50 frames.
Now suppose, the same OCR model processes the same 10 second feed  at 10 FPS instead 
of 5, i.e., this model processes a total of 100 frames.
If this setting results in more useful frames, then running at a
higher frame rate would lead to shorter scanning times on average.

To study this, we analyze videos of users scanning cards from our user study
described in Section ~\ref{sec:eval:user_study}. We \emph{simulate} different
frame rates by extracting frames at differently spaced intervals from the recorded
videos. Closer intervals represent faster frame rates and possibly identical frames,
and vice versa for wider intervals.  We then pass these frames through two different
OCR models (Boxer OCR and \system OCR) and for each frame rate we compute the percentage
of useful frames obtained to the total number of frames processed.

\begin{figure}[t]
  \centering
  \includegraphics[width=1\columnwidth]{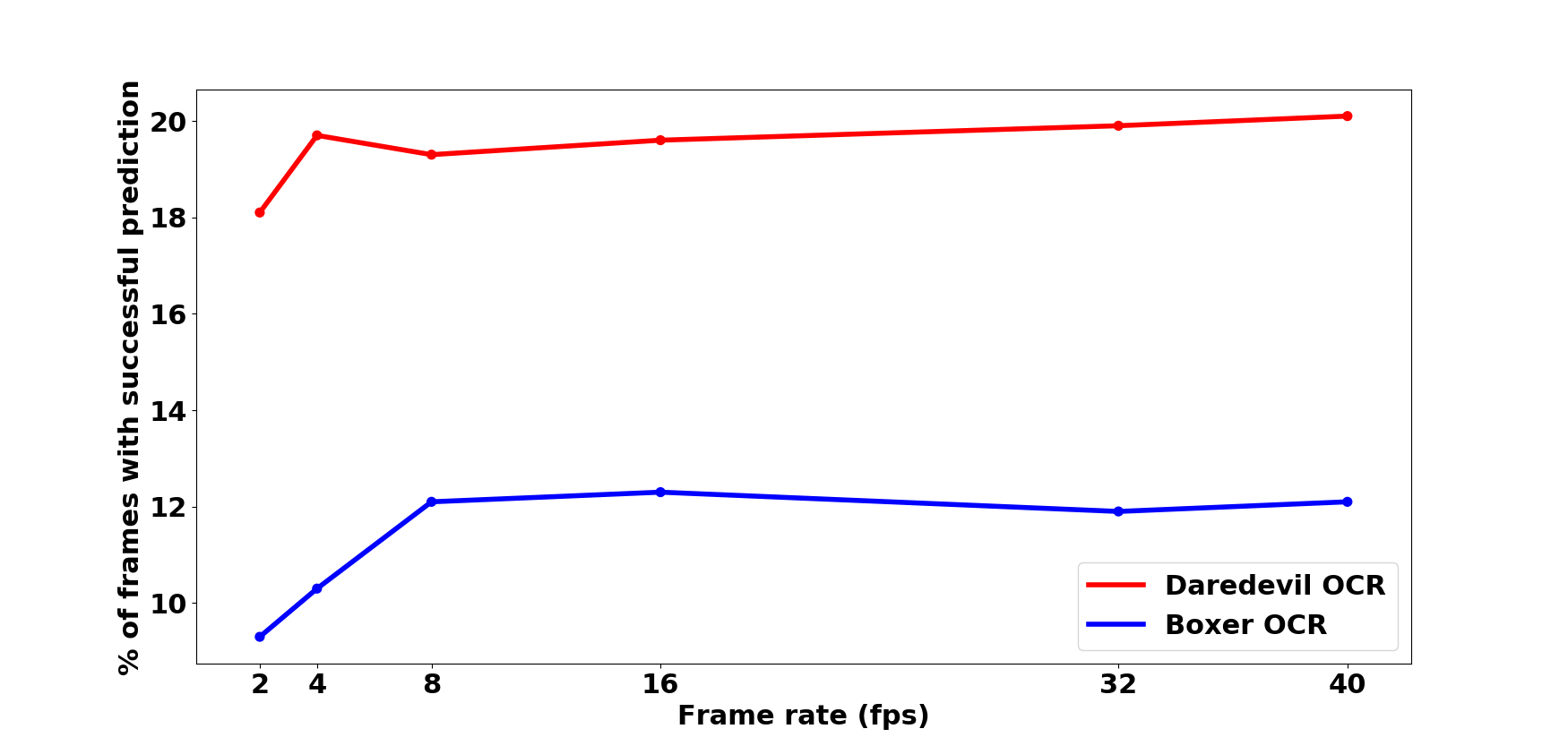}
  \caption{Plot shows that as the frame rates increase, the fraction of frames with 
    successful predictions roughly remains constant, meaning that the number of frames
    with successful predictions increases with frame rate. Thus, systems enhancements
    to increase the frame rate, even with the same machine learning model can
    lead to faster scanning times.}
  \label{fig:pred_ratio_vs_fps}
  \hrulefill
\end{figure}

Figure \ref{fig:pred_ratio_vs_fps} plots the variation of frame rates
to percentage of frames with successful predictions averaged over 27
different scanning videos sampled from our user study. The plots are roughly 
constant for both Boxer OCR and \system OCR. This indicates that with increasing 
frame rates and  correspondingly increasing the number of frames processed
by the  models, the number of useful frames (i.e., the number of frames
on which we are successfully able to run OCR) also increases. These results
suggest that even closely spaced frames contain sufficient diversity leading
to different, and possibly correct predictions with the same machine learning
model. Thus, further systems enhancements that lead to higher frame rates with
the same OCR model contribute to faster scan times and better user experience.

\bibliographystyle{plain}
\bibliography{daredevil.bib}

%%%%%%%%%%%%%%%%%%%%%%%%%%%%%%%%%%%%%%%%%%%%%%%%%%%%%%%%%%%%%%%%%%%%%%%%%%%%%%%%
\end{document}